\documentclass[useAMS,usenatbib]{mn2e}

\usepackage{graphicx}
\usepackage{amsmath}
\usepackage{amssymb}
\usepackage{color}
\usepackage{subfig}
\usepackage[breaklinks,colorlinks,citecolor=blue,linkcolor=red]{hyperref} 
\usepackage[all]{hypcap}

\newcommand\msun{\, \rm M_\odot}

\newcommand\kms{\, \rm km\,s^{-1}}

\newcommand\eout{{e_{\rm out}}}
\newcommand\ein{{e_{\rm in}}}
\newcommand\aout{{a_{\rm out}}}
\newcommand\ain{{a_{\rm in}}}
\newcommand\amax{{a_{\rm 3}^{\rm max}}}
\newcommand\mbh{{M_{\rm BH}}}
\newcommand\mwd{{M_{\rm WD}}}

\newcommand\sigbh{{\sigma_{\rm BH}}}

\newcommand\sigwd{{\sigma_{\rm WD}}}

\newcommand\vk{{v_\mathrm{k}}}
\newcommand{\be}{\begin{equation}}
\newcommand{\ee}{\end{equation}}

\title[WD disruptions by BHs in triples]{Electromagnetic transients and gravitational waves from white dwarf disruptions by stellar black holes in triple systems}
\author[G. Fragione et al.]{  \parbox{\textwidth}{Giacomo Fragione$^{1,2,3}$\thanks{E-mail: giacomo.fragione@northwestern.edu}, Brian D. Metzger$^{4}$, Rosalba Perna$^{5,6}$, Nathan W. C. Leigh$^{7,8}$, Bence Kocsis$^{9}$\vspace*{0.3cm}}\\
$^1$Department of Physics \& Astronomy, Northwestern University, Evanston, IL 60202, USA\\
$^2$Center for Interdisciplinary Exploration \& Research in Astrophysics (CIERA), Evanston, IL 60202, USA\\
$^3$Racah Institute for Physics, The Hebrew University, Jerusalem 91904, Israel\\
$^4$Department of Physics, Columbia University, New York, NY 10027, USA\\
$^5$Department of Physics and Astronomy, Stony Brook University, Stony Brook, NY 11794-3800, USA\\
$^6$Center for Computational Astrophysics, Flatiron Institute,  New York, NY 10010, USA\\
$^7$Departamento de Astronom\'ia, Facultad de Ciencias F\'isicas y Matem\'aticas, Universidad de Concepci\'on, Concepci\'on, Chile\\
$^8$Department of Astrophysics, American Museum of Natural History, New York, NY 10024, USA\\
$^9$Institute of Physics, E\"{o}tv\"{o}s University, P\'azm\'{a}ny P. s. 1/A, Budapest, 1117, Hungary
}

\hypersetup{draft}

\begin{document}

\maketitle

\begin{abstract}
Mergers of binaries comprised of compact objects can give rise to explosive transient events, heralding the birth of exotic objects which cannot be formed through single star evolution. Using a large number of direct N-body simulations, we explore the possibility that a white dwarf (WD) is dynamically driven to tidal disruption by a stellar-mass black hole (BH) as a consequence of the joint effects of gravitational wave (GW) emission and Lidov-Kozai oscillations imposed by the tidal field of a outer tertiary companion orbiting the inner BH-WD binary. We explore the sensitivity of our results to the distributions of natal kick velocities imparted to the BH and WD upon formation, adiabatic mass loss, semi-major axes and eccentricities of the triples, and stellar mass ratios. We find rates of WD-TDEs in the range $1.2\times 10^{-3}-1.4$ Gpc$^{-3}$ yr$^{-1}$ for $z\leq 0.1$, rarer than stellar TDEs in triples by a factor of $\sim 3$--$30$. The uncertainty in the TDE rates may be greatly reduced in the future using gravitational wave (GW) observations of Galactic binaries and triples with LISA. WD-TDEs may give rise to high energy X-ray or gamma-ray transients of duration similar to long gamma-ray bursts but lacking the signatures of a core-collapse supernova, while being accompanied by a supernova-like optical transient which lasts for only days. WD--BH and WD--NS binaries will also emit GWs in the LISA band before the TDE. The discovery and identification of triple-induced WD-TDE events by future time domain surveys and/or GWs could enable the study of the demographics of BHs in nearby galaxies. 
\end{abstract}


\section{Introduction}

Mergers of binaries comprised of two compact objects have been the subject of numerous theoretical investigations over the past several years. This interest is motivated in part by the fact that such binaries give rise to explosive and potentially luminous transient events, which leave behind exotic objects that cannot otherwise form from single stars at the end of their lifetimes.  The coalescence of binary stellar black holes (BHs) and neutron stars (NSs) have been observed by LIGO-Virgo via their gravitational wave (GW) emission \citep{ligo2018}.  Thanks to the discovery of gamma-ray and non-thermal afterglow emission in coincidence with the LIGO-detected merger GW170817 \citep{Abbott2017NS}, binary NS mergers are now the confirmed progenitor of at least one short gamma-ray burst (GRB).  NS-BH mergers may also produce short GRBs, at least for small mass ratios and high BH spin such that the NS is tidally disrupted outside of the BH horizon (e.g.~\citealt{Foucart2018}). The coalescence of binary white dwarfs (WDs) provide a likely pathway to produce Type Ia supernovae \citep[SNe;][]{katz2012,maoz2014,livio2018,toonen2018,hamers2018}.

Mergers of NS-WD and BH-WD binaries are expected to occur as well. To date, $\sim 20$ NS-WD binaries have been confirmed in our Galaxy \citep{vank2005}, while only one field BH-WD binary candidate is presently known \citep{Bahramian+17}.  In such binaries, the WD may approach the NS or BH close enough to be disrupted as a tidal disruption event (WD-TDE).  For example, such coalescence events could result from GW emission in isolation \citep{Metzger12} or as a consequence of non-coherent scatterings in star clusters \citep{leigh14,Kremer2019}.

What makes the mergers of NS-WD and BH-WD binaries of particular interest is the possibility that they could generate peculiar transients. Several works have characterized the possible electromagnetic (EM) signatures of the tidal disruption of a WD by a NS or a BH \citep{Fryer+99,King+07,Metzger12,Fernandez&Metzger13,Margalit&Metzger16,bob2017,toonen2018,Zenati+19,Fernandez+19}. In particular, such events may produce a high energy transient similar to a GRB, or thermal emission similar to a short-lived supernova \citep{Metzger12,Zenati+19}. Interesting is also the case in which the WD is disrupted by an intermediate-mass black hole (IMBH; \citealt{rosswog2008,Macleod+16,frlgk2018}); tidal compression during such WD-IMBH events could generate a large quantity of $^{56}$Ni capable of powering a peculiar Type Ia-like supernova. NS-WD, BH-WD, and IMBH-WD encounters also produce GW emission up to the point of disruption, observable by the planned \textit{LISA} detector.

In this paper, given the preponderance of triples in the Galaxy \citep[e.g.][]{leigh13}, we explore a new triple channel of WD-BH merger events, in which the WD is driven sufficiently close to the BH to be tidally disrupted as a consequence of the joint effect of GW emission and Lidov-Kozai (LK) evolution imposed by the tidal field of a third companion that orbits the BH-WD binary.  We start from the progenitors of the BH and WD and model the effects of natal kicks during the formation of the compact objects (e.g. BH birth in a supernova) and the survival of the triples. Given the many uncertainties involved in the modelling of binary evolution, we explore a variety of models which make different assumptions about the distributions of natal kicks, semi-major axes and eccentricities of the triple, and initial stellar mass ratios. In total, we run $\sim 10^{4}$ direct $N$-body simulations to explore the prospects for BH-WD systems. We determine how the probability of a WD-TDE depends on these assumptions, map the parameter distributions of merging systems back to the initial distributions, and compute the WD-TDE rate in the Universe through the triple channel.

The paper is organized as follows. In Section~\ref{sect:init}, we present our numerical methods and describe the properties of the triple population that we evolve. In Section~\ref{sect:results}, we discuss the parameters of the merging systems. The implications for possible electromagnetic counterparts are presented in Section~\ref{sect:emcount}, and for gravitational waves in Section~5. Finally, in Section~\ref{sect:conc}, we discuss the implications of our findings and draw our conclusions.

\section{Triple population}
\label{sect:init}

\begin{table*}
\caption{Models: name, mean of WD kick-velocity distribution ($\sigma$), eccentricity distribution ($f(e)$), maximum outer semi-major axis of the stellar progenitor triple ($\amax$), fraction of stable systems after SNe ($f_{\rm stable}$), fraction of WD-TDEs from the $N$-body simulations ($f_{\rm WD-TDE}$), fraction of stable systems assuming adiabatic mass loss ($f_{\rm stable}^{\rm (ad)}$).}
\centering
\begin{tabular}{lccccccccc}
\hline
Name & $\sigbh$ ($\kms$) & $\sigwd$ ($\kms$) & $f(q)$ & $f(a)$ & $f(e)$ & $\amax$ (AU) & $f_{\rm stable}$ & $f_{\rm WD-TDE}$ & $f_{\rm stable}^{\rm (ad)}$ \\
\hline\hline
A1 & $34$ & $0.5$ & uniform & log-uniform & uniform  & $5000$ & $7.2\times 10^{-5}$ & $0.21$ & $4.6\times 10^{-4}$\\
A2 & $13$ & $0.5$ & uniform & log-uniform & uniform  & $5000$ & $5.2\times 10^{-4}$ & $0.15$ & $5.1\times 10^{-3}$\\
A3 & $0$  & $0.5$ & uniform & log-uniform & uniform  & $5000$ & $1.2\times 10^{-2}$ & $0.16$ & $2.4\times 10^{-1}$\\
A4 & $0$  & $0$   & uniform & log-uniform & uniform  & $5000$ & $1.1\times 10^{-2}$ & $0.14$ & $2.1\times 10^{-1}$\\
B1 & $34$ & $0.5$ & uniform & uniform     & uniform  & $5000$ & $1.6\times 10^{-5}$ & $0.19$ & $4.9\times 10^{-5}$\\
C1 & $34$ & $0.5$ & uniform & log-uniform & thermal  & $5000$ & $1.0\times 10^{-4}$ & $0.20$ & $4.7\times 10^{-4}$\\
D1 & $34$ & $0.5$ & uniform & log-uniform & uniform  & $2000$ & $1.0\times 10^{-4}$ & $0.28$ & $6.6\times 10^{-4}$\\
D2 & $34$ & $0.5$ & uniform & log-uniform & uniform  & $7000$ & $7.3\times 10^{-5}$ & $0.24$ & $5.0\times 10^{-4}$\\
E1 & $34$ & $0.5$ & -       & log-uniform & uniform  & $5000$ & $1.7\times 10^{-3}$ & $0.31$ & $9.4\times 10^{-3}$\\
\hline
\end{tabular}
\label{tab:models}
\end{table*}

General schemes for population synthesis in triples have been developed by a number of authors \citep{kratt2012,toonen2016,toon2018a,toon2018b}. The stellar triples in our simulations are initialized as follows. In total, we consider nine different sets of initial conditions (see Table~\ref{tab:models}).

In all models, we adopt the \citet{kroupa2001} initial mass function in the relevant mass range
\begin{equation}
f(m) = 0.0795\,\msun^{-1} (m/\msun)^{-2.3}\quad{\rm if~} m \geq 0.5\msun
\label{eqn:bhmassfunc}
\end{equation}
where the constant coefficient takes into account the fraction of stars with $m< 0.5\msun$ such that the integral of $\int_0^{\infty} f(m) dm = 1$. We draw the stellar progenitor of the most massive star in the inner binary $m_1$ from the mass range $20\msun$--$150\msun$, which we assume collapses to a BH. The exact value of the BH mass depends on details of the stellar evolution related to, for example, metallicity, stellar winds and rotation. However, for simplicity, we assume that $\mbh=m_1/3$ \citep{sil17,fff2019}. In our \textit{fiducial} model, we adopt a flat mass ratio distribution for both the inner and outer orbit \citep{sana12,duch2013}.\footnote{Note that the this is different from \citet{sil17} who assumed a log-uniform mass ratio distribution, $f(q)\propto q^{-1}$. \citet{duch2013} find that $f(q)\propto q^{1.16\pm0.16}$ and $q^{-0.01\pm0.03}$ for solar type stars with period less than or larger than $10^{5.5}\,$day, respectively, while \citet{sana13} find $f(q)\propto q^{-1.0\pm 0.4}$ for massive O-type stars.} 
The mass of the secondary in the inner binary is sampled within the range $1$--$8\msun$. We assume that this star gives birth to a WD of mass \citep{hurley00}
\begin{equation}
\mwd=0.109\ m_2+0.394
\end{equation}
and radius
\begin{equation}\label{eq:Rwd}
R_{\rm WD}=\max\left[10\ \mathrm{km},0.01\ \mathrm{R}_\odot\sqrt{\left(\frac{M_{\rm ch}}{\mwd}\right)^{2/3}-\left(\frac{\mwd}{M_{\rm ch}}\right)^{2/3}}\right]\ ,
\end{equation}
where $M_{\rm ch}=1.44\msun$ is the Chandrasekhar mass. The mass of the third companion ($m_3$) is drawn from the range $0.5$-$150\msun$. We note that we assume that if the mass of the tertiary is in the range $1$-$8\msun$ it generates a WD, if in the range $8 \msun$--$20\msun$ it collapses to a NS of mass $1.3\msun$, and if in the range $20 \msun$--$150\msun$ it collapses to a BH of mass $m_3/3$ \citep{sil17}. We run one model where all the masses are drawn independently from each other from Eq.~\ref{eqn:bhmassfunc}. For comparison, we also estimate how the final WD-TDE rate changes if the mass ratio distribution is assumed to be log-uniform \citep{sana13}. The distributions of the inner and outer semi-major axes, $\ain$ and $\aout$ (respectively), are assumed to be log-uniform \citep{kob2014}, but we also consider a model with uniform distributions of inner and outer semi-major axes. Other alternatives would include log-normal and other power-laws \citep{moe2017,igos2019}. We set as a minimum initial orbital separation $\ain(1-e_{\rm in}^2)\approx 10$ AU to avoid mass transfer \citep[e.g.][]{ant17}, and adopt different values for the initial maximum separation of the triple $\amax=2000$ AU--$5000$ AU--$7000$ AU \citep{sana2014}. For what concerns the orbital eccentricities $\ein$ and $\eout$, we assume flat distributions \citep[e.g.][]{geller19}. For comparison, we run one additional model where we take a thermal distribution of eccentricities. Finally, the initial mutual inclination $i$ between the inner and outer orbits is drawn from an isotropic distribution, while the other relevant angles are drawn from uniform distributions\footnote{\citet{tok2017} has shown that low-mass triples with separations $\lesssim 1000$ AU have a much flatter configuration.}.

After sampling the relevant parameters, we check that the initial configuration satisfies the stability criterion of \citet{mar01} for stable hierarchical triples. If this is not the case, we sample again the triple parameters as explained above. Otherwise, we let the primary star in the inner binary undergo an SN explosion and instantaneously convert it to a BH. We note that in reality not all the mass is ejected during the SN process, but part of it can be lost previously through stellar winds \citep{kratt2012,michp2014,michper2019}. As a consequence, the Blauuw kick due to mass loss would be typically smaller, thus possibly unbinding a smaller number of triples. As a result of the mass loss, the exploding star is imparted a kick to its center of mass \citep{bla1961}, and the system receives a natal kick due to recoil from an asymmetric supernova explosion. We assume that the BH natal velocity kick is drawn from a Maxwellian distribution
\begin{equation}
p(\vk)\propto \vk^2 e^{-\vk^2/\sigma^2}\ ,
\label{eqn:vkick}
\end{equation}
with a mean velocity $\sigma$. We implement momentum-conserving kicks, in which we assume that the momentum imparted to a BH is the same as the momentum given to a NS \citep{fryer2001}. As a consequence, the kick velocities for the BHs are lowered by a factor of $1.4\msun/\mbh$ with respect to those of NSs. The value of $\sigma$ is highly uncertain. We adopt $\sigma=260 \kms$ for NSs, consistent with the distribution deduced by \citet{hobbs2005}, but we also run an additional model where we set $\sigma=100 \kms$, consistent with the distribution of natal kicks found by \citet{arz2002}. We also explore a model where no natal kick is imparted during BH formation.

We update the orbital elements of the triple as appropriate \citep{pijloo2012,lu2019,fff2019}, checking that the new configuration satisfies the stability criterion for stable hierarchical triples \citep{mar01}. If the system remains stable, we assume that the secondary forms a WD. Also in this case, we assume that the WD natal velocity kick is drawn from a Maxwellian distribution. We consider both models in which the WDs receive no natal velocity kick ($\sigwd=0\kms$) and models where the mean velocity is $\sigwd=0.5\kms$. This is consistent with the new findings of \citet{elb2018}, who found breaks in the separation distribution of MS-WD and WD-WD binaries in \textit{Gaia} data, which could be explained if WDs incur a natal velocity kick at their formation\footnote{This could reflect the mass-loss in the intermediate regime between prompt and adiabatic mass-loss, rather than a natal kick.}, with a magnitude of $\sim 0.75\kms$. After the second SN event, we update again the orbital elements of the triple and again check that it is stable. Finally, if the third companion is more massive than $1\msun$, we let it undergo conversion into either a WD, NS or BH, of mass $m_3^{\rm fin}=0.109\ m_3+0.394$, $m_3^{\rm fin}=1.3\msun$, $m_3^{\rm fin}=m_3/3$, if $1\msun \le m_3< 8\msun$, $8\msun \le m_3< 20\msun$, $m_3> 20\msun$, respectively. If $m_3<1$, then $m_3^{\rm fin}=m_3$. Many systems turn out to occupy a quasi-secular regime, for which the behavior is somewhat different and more chaotic than the secular LK mechanism \citep{antoper12,fragrish2018}.

Table~\ref{tab:models} reports the fraction of systems that are stable after all the SNe have taken place; this is denoted by $f_{\rm stable}$ for each of our models.

We integrate the triple systems by means of the \textsc{ARCHAIN} code \citep{mik06,mik08}, including PN corrections up to order PN2.5. We perform $\sim 1000$ simulations for each model in Table~\ref{tab:models}, and impose a number of stopping conditions  as follows:
\begin{itemize}
\item The system undergoes $1000$ LK-cycles, i.e. the total time exceeds $1000$ $T_{\rm LK}$, where the triple LK timescale is
\begin{equation}
T_{\rm LK}=\frac{8}{15\pi}\frac{m_{\rm tot}}{m_3^{\rm fin}}\frac{P_{\rm out}^2}{P_{\rm in}}\left(1-e_{\rm out}^2\right)^{3/2}\ ,
\end{equation}
where $m_{\rm tot}=\mbh+\mwd+m_3^{\rm fin}$ and $P_{\rm in}$ and $P_{\rm out}$ are the inner and outer orbital periods, respectively. 
\item The WD is tidally disrupted by the BH in the inner binary due to a high orbital eccentricity. This occurs whenever their relative distance becomes smaller than the tidal disruption radius,
\begin{equation}
R_{\rm t}=R_{\rm WD} \left(\frac{\mbh}{\mwd}\right)^{1/3}\ .
\label{eqn:rtid}
\end{equation}
\item The system age exceeds 10 Gyr.
\end{itemize}

\section{Results}
\label{sect:results}

\begin{figure} 
\centering
\includegraphics[scale=0.55]{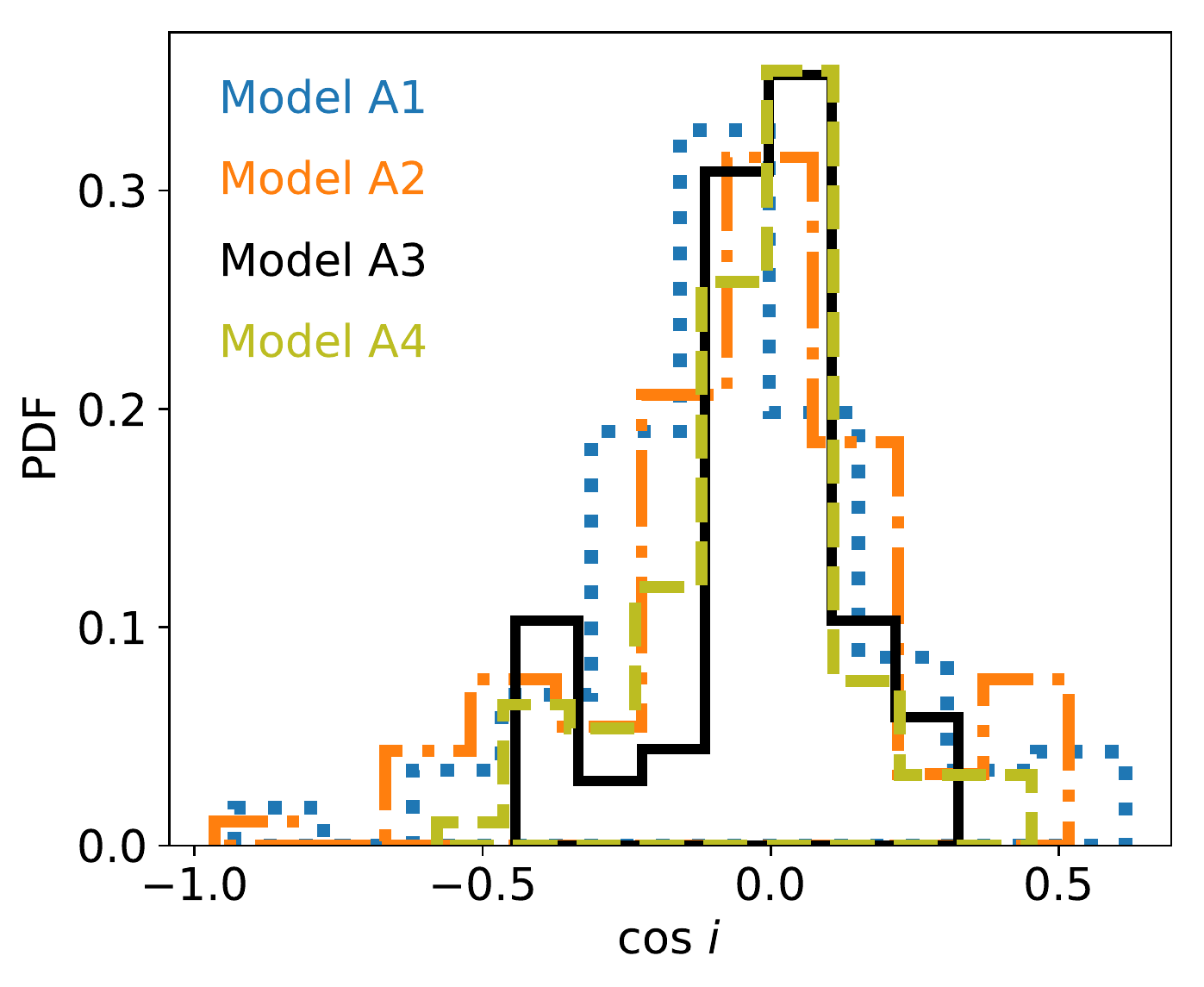}
\caption{Inclination PDF of merging BH-MS binaries in triples, for $\amax=5000$ AU and different values of the mean kick velocity $\sigbh$ and $\sigwd$ (Models A1-A4).}
\label{fig:incl}
\end{figure}

\begin{figure} 
\centering
\includegraphics[scale=0.55]{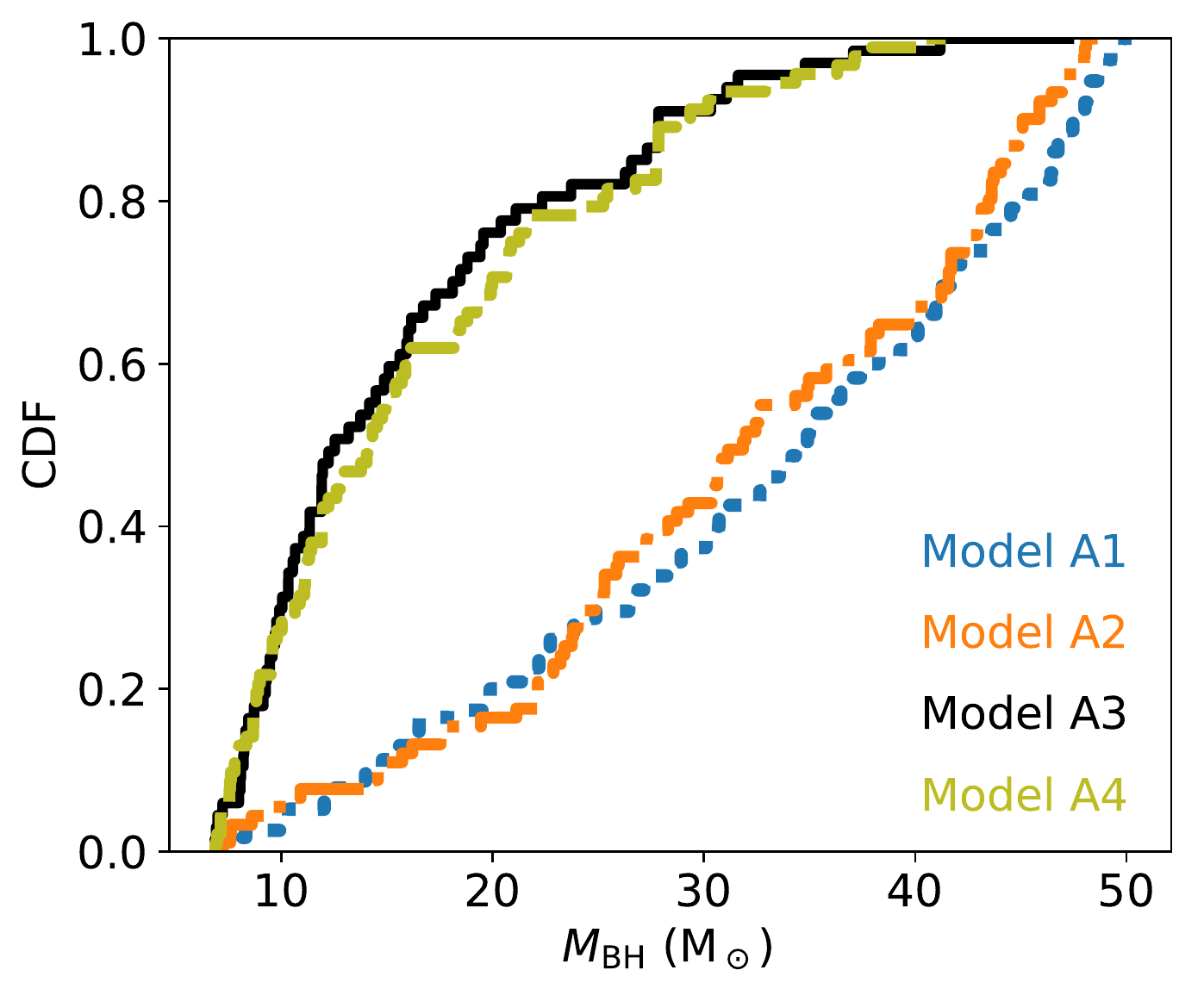}
\includegraphics[scale=0.55]{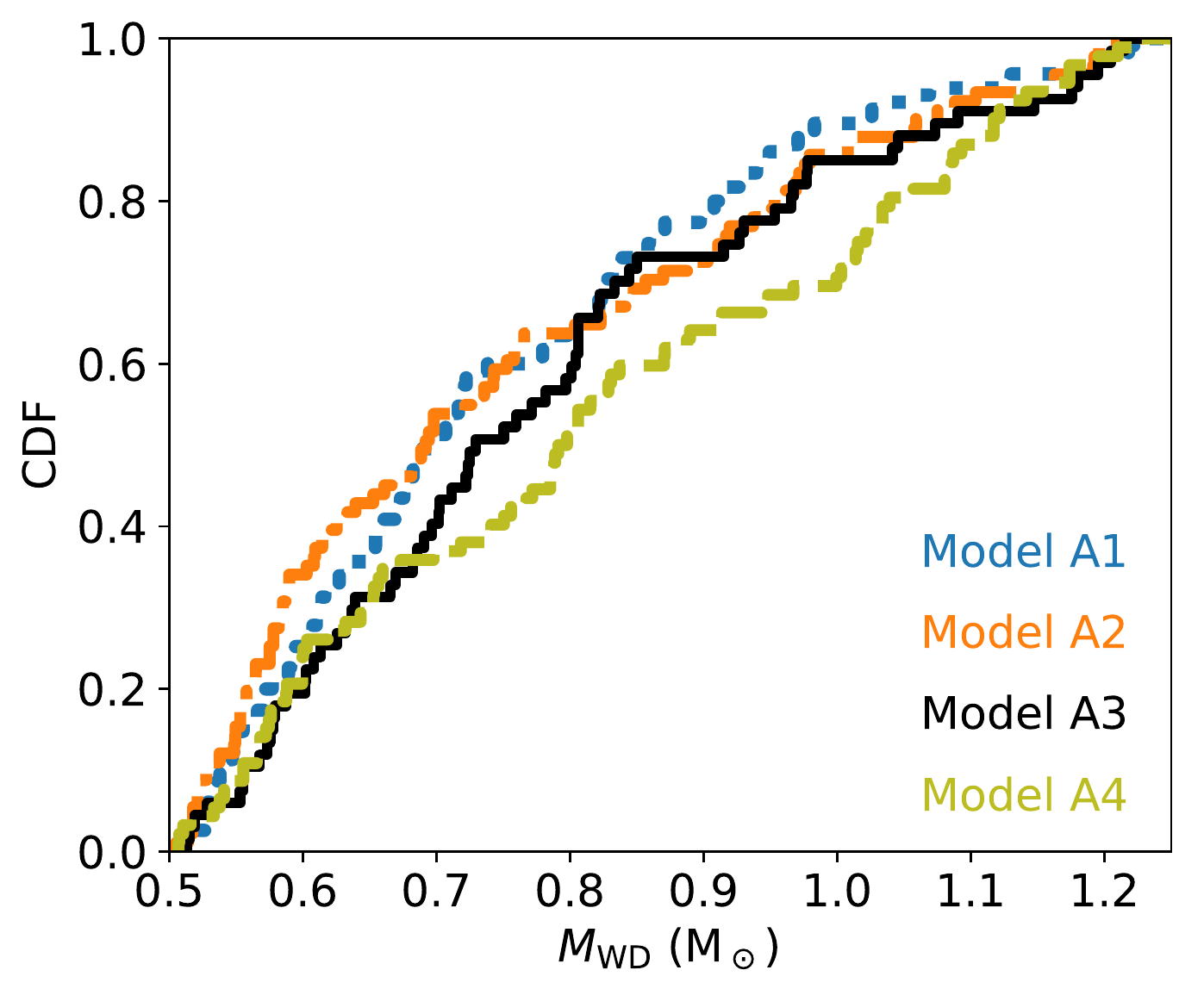}
\caption{Cumulative distribution function of the mass of BHs (top) and WDs (bottom) in BH-WD binaries in triples that lead to a WD-TDE, for different values of $\sigbh$ and $\sigwd$ (Models A1-A4).}
\label{fig:mbhwd}
\end{figure}

\subsection{Inclination}

A BH-WD binary is expected to be significantly perturbed by the tidal field of the third companion whenever its orbital plane is sufficiently inclined with respect to the outer orbit \citep{lid62,koz62}. Figure~\ref{fig:incl} shows the inclination probability distribution function (PDF) of systems that lead to a WD-TDE. We show the distributions for $\amax=5000$ AU and different values of $\sigbh$ and $\sigwd$, Models A1--A4 (see Table~\ref{tab:models}). Most of the WD-TDEs in triples occur when the inclination approaches $\sim 90^\circ$, since in this case the LK mechanism is efficient at pumping $\ein$ up to unity.

\begin{figure*} 
\centering
\includegraphics[scale=0.55]{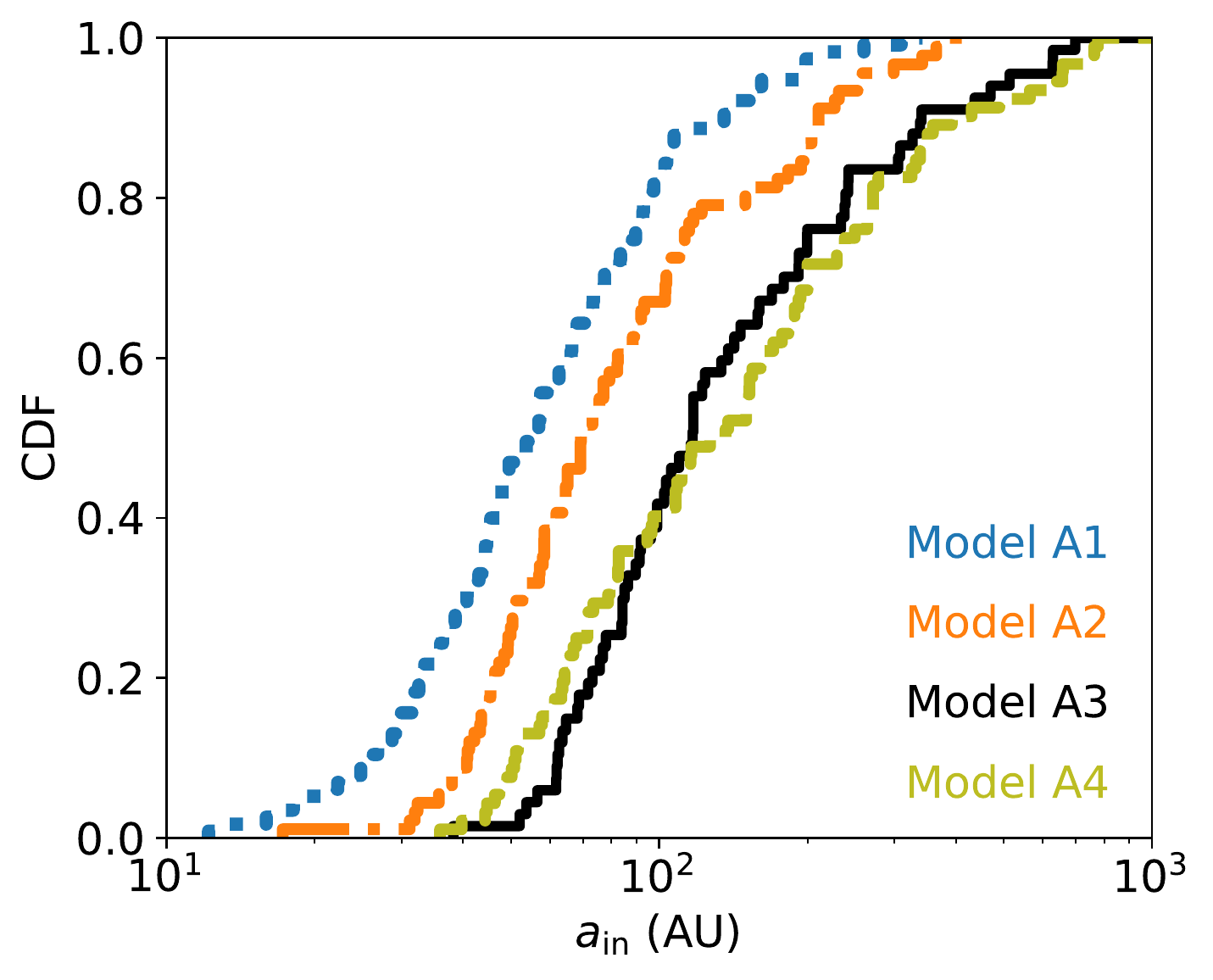}
\hspace{1cm}
\includegraphics[scale=0.55]{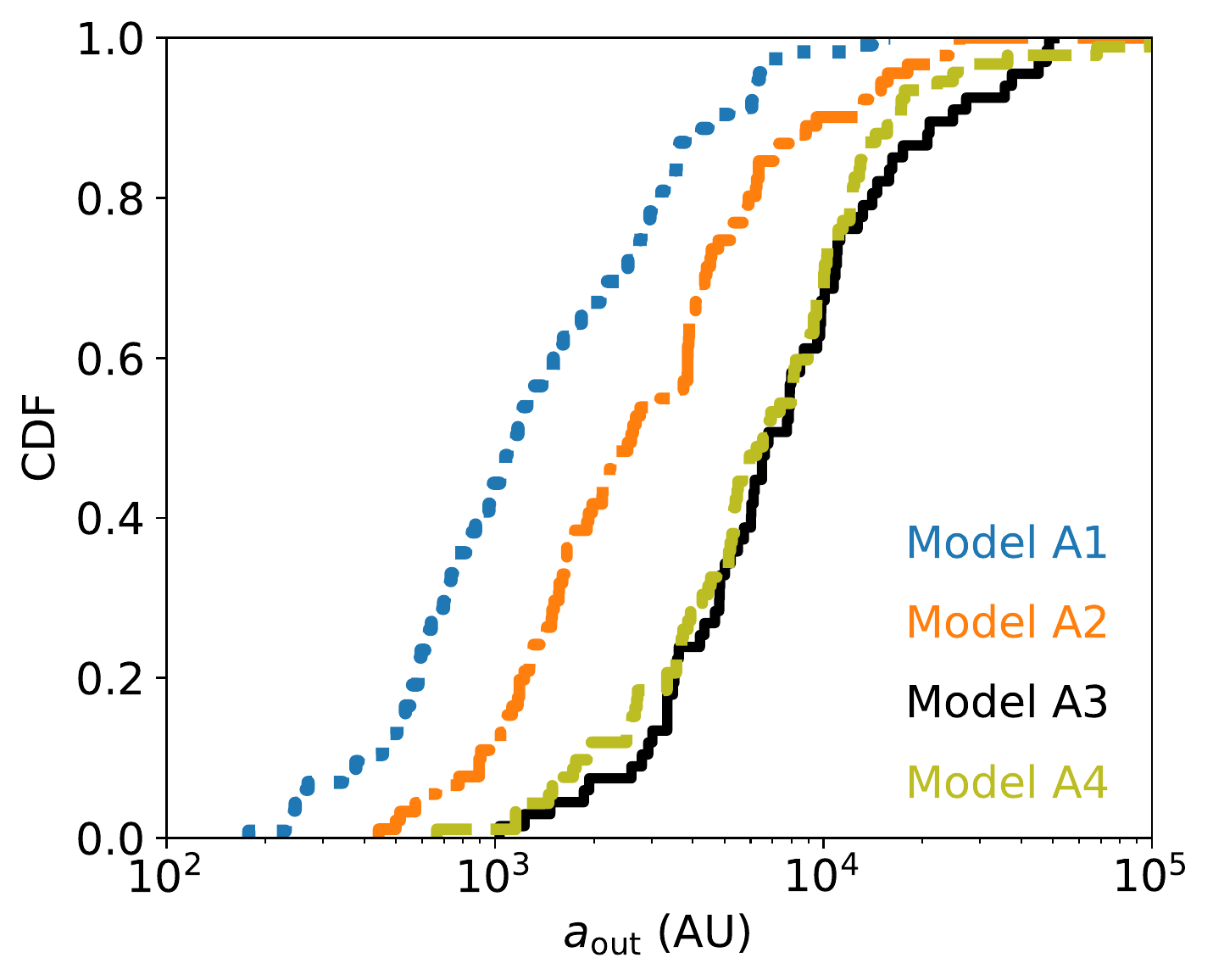}\\
\includegraphics[scale=0.55]{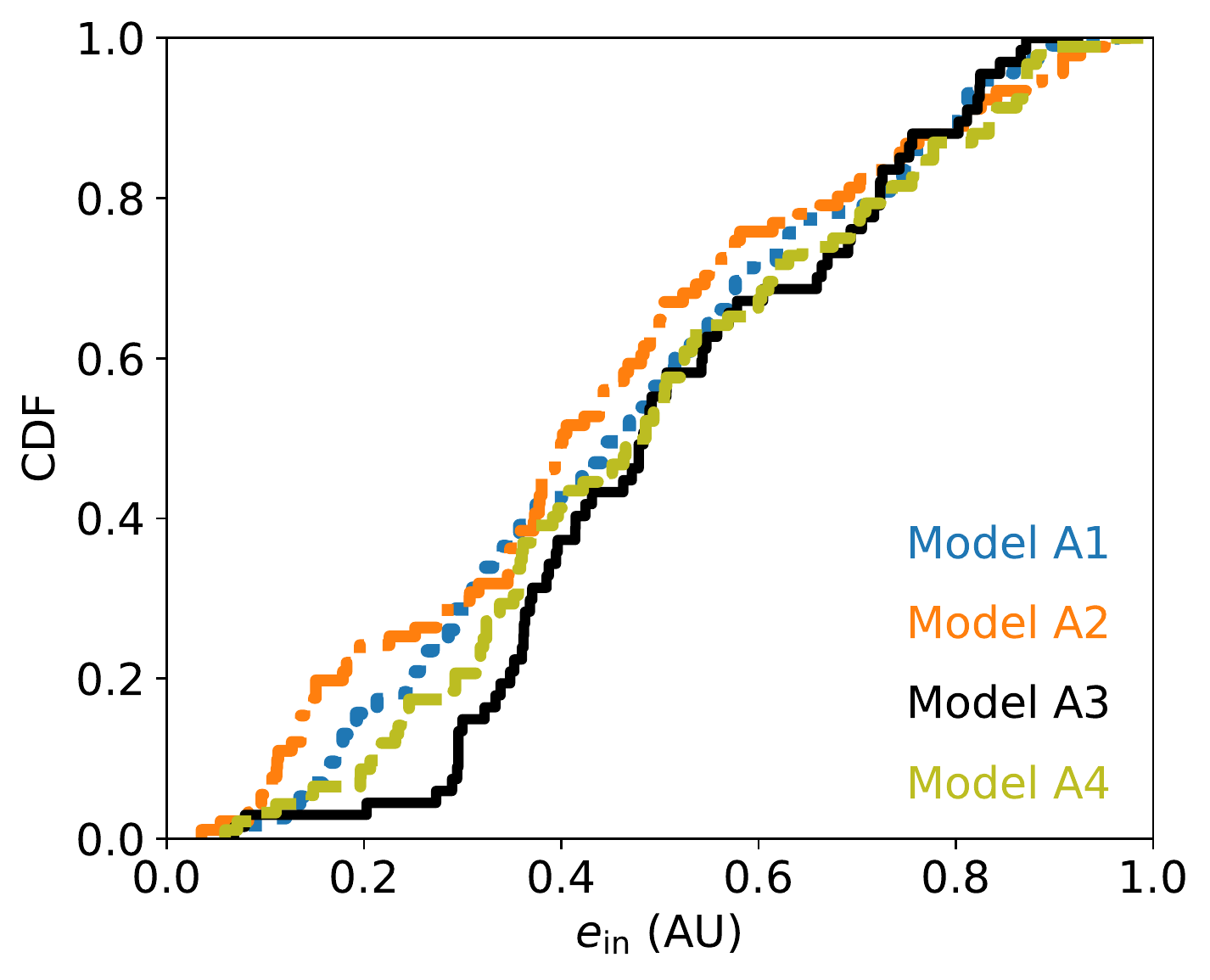}
\hspace{1cm}
\includegraphics[scale=0.55]{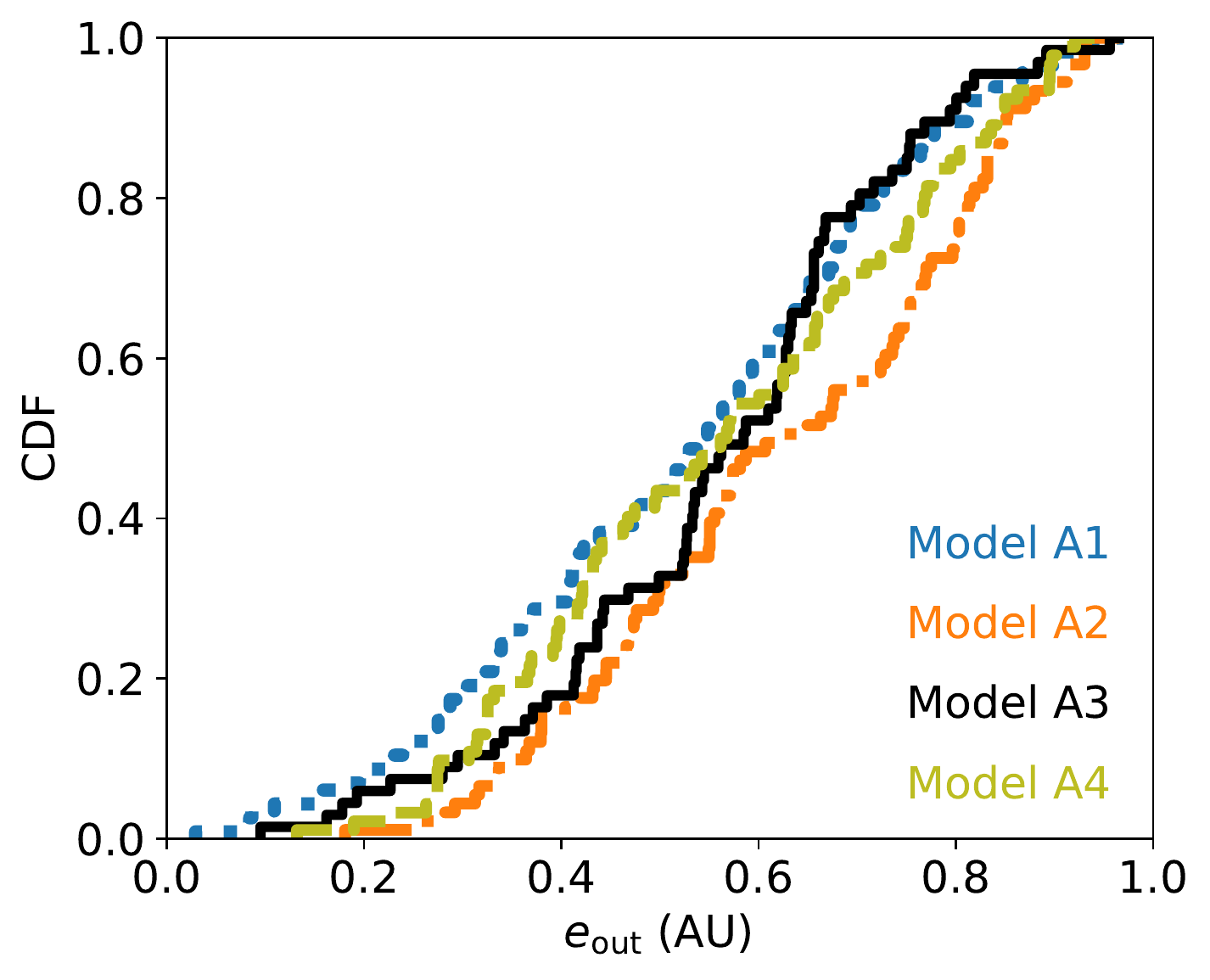}\\
\caption{Cumulative distribution function of the initial inner (left) and outer (right) semi-major axes (top) and eccentricities (bottom) of BH-WD binaries in triples that lead to a WD-TDE, for different values of $\sigbh$ and $\sigwd$ (Models A1-A4). }
\label{fig:ainaout}
\end{figure*}

\subsection{Mass of black hole and white dwarf}

Figure~\ref{fig:mbhwd} shows the cumulative distribution function (CDF) of $\mbh$ (top panel) and $\mwd$ (bottom panel) for systems that produce a WD-TDE for Models A1-A4. Systems with high values of $\sigbh$ prefer higher BH masses. This is explained by our assumption of momentum-conserving kicks, where BHs receive a kick scaled by $1/\mbh$. Thus, more massive BHs are imparted lower velocity kicks on average and are more likely to be retained in bound triples, which eventually produce a WD-TDE. The distribution of the mass of the WDs does not display a strong dependence on the assumed mean velocity kicks for BHs and WDs.

\subsection{Inner and outer semi-major axes}

The choice of $\sigbh$ affects the distribution of the orbital parameters of BH-WD systems that lead to a WD-TDE. Figure~\ref{fig:ainaout} shows the CDF of the inner (left) and outer (right) semi-major axes (top) and eccentricities (bottom) of BH-WD binaries in triples that lead to a WD-TDE, for different values of $\sigbh$ and $\sigwd$. As also shown in \citet{fff2019}, we find that larger mean natal kicks imply smaller inner and outer semi-major axes. This is because high velocity kicks preferentially unbind triple systems with wide orbits. The inner and outer eccentricities, however, do not depend on the assumed value of $\sigbh$. Also, the value of $\sigwd$ does not affect the distribution of the orbital elements of systems that produce a WD-TDE.

\begin{figure} 
\centering
\includegraphics[scale=0.55]{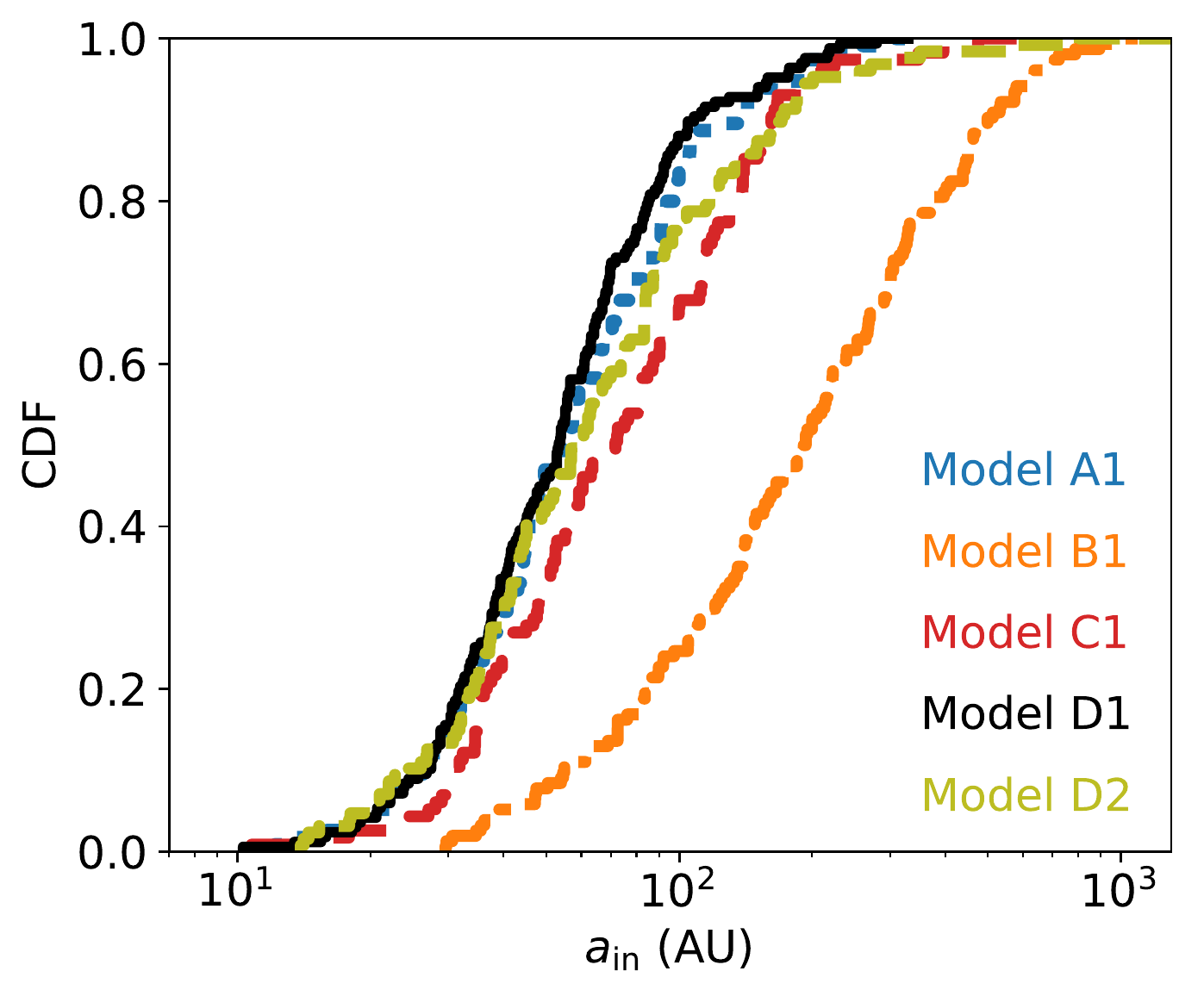}
\includegraphics[scale=0.55]{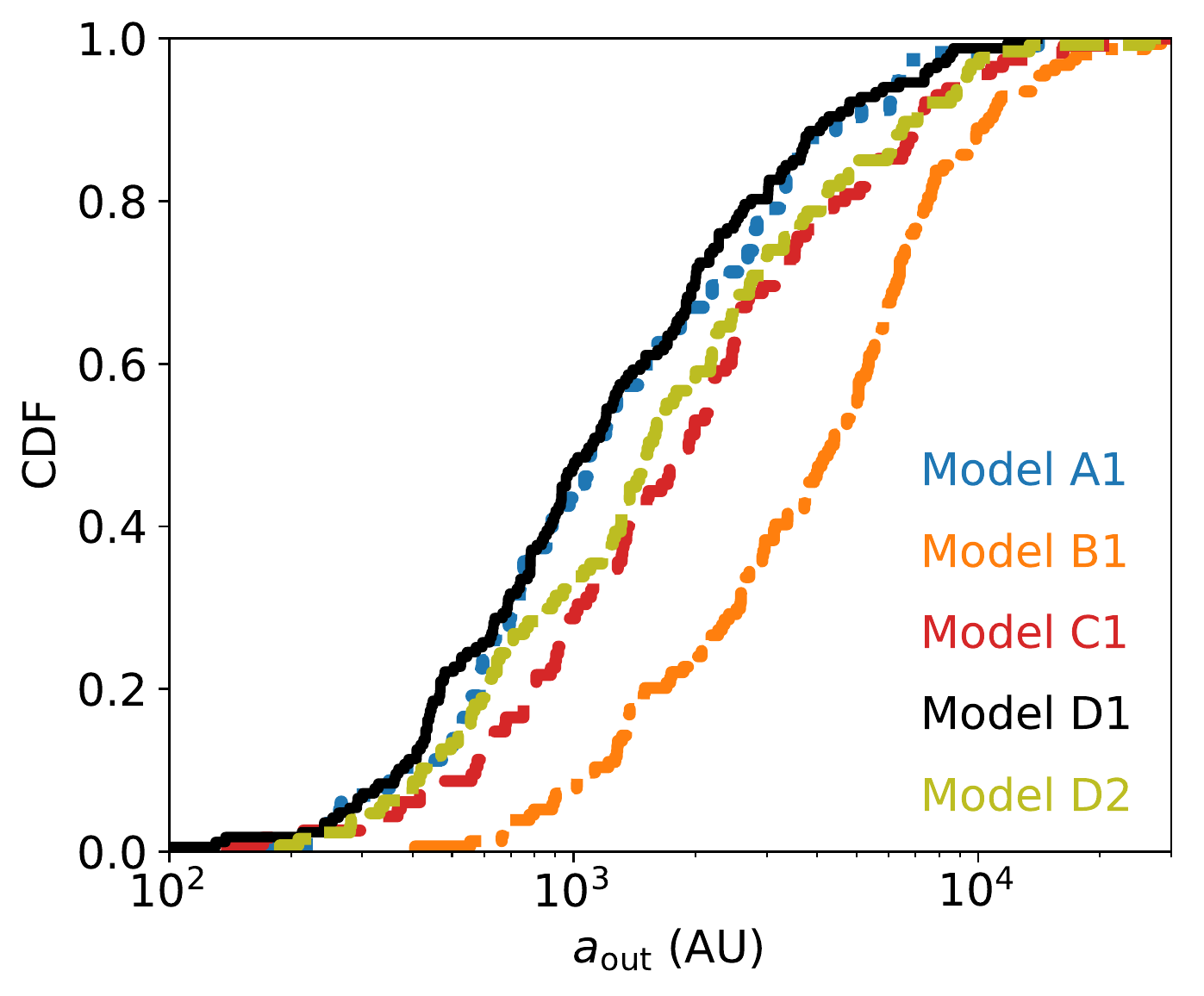}
\caption{Cumulative distribution function of the inner (left) and outer (right) semi-major axes (top) and eccentricities (bottom) of BH-WD binaries in triples that lead to a WD-TDE, for different initial distributions of semi-major axes and eccentricities.}
\label{fig:ainaout2}
\end{figure}

Fig.~\ref{fig:ainaout2} shows how the distributions of $\ain$ and $\aout$ of BH-WD systems that lead to a WD-TDE depend on the initial distribution of the orbital elements and $\amax$. We find that larger values of $\amax$ lead to larger inner and outer semi-major axes, though the dependence on this parameter is not significant.  Model C1, where an initial thermal distribution of $\ein$ and $\eout$ is assumed, predicts a distribution similar to Model D2, where $\amax=7000$ AU. The CDFs are significantly affected by the choice of the initial distribution for $\ain$ and $\aout$. We find that $\sim 50\%$ of the BH-WD systems that lead to a WD-TDE have $\ain\lesssim 50$ AU and $\aout\lesssim 1000$ AU in Model A1 ($f(a)$ log-uniform) and $\ain\lesssim 200$ AU and $\aout\lesssim 5000$ AU in Model B1 ($f(a)$ uniform). Also in this case, the distributions for $\ein$ and $\eout$ do not depend on the details of the initial conditions.

\subsection{Rates}

\begin{figure} 
\centering
\includegraphics[scale=0.55]{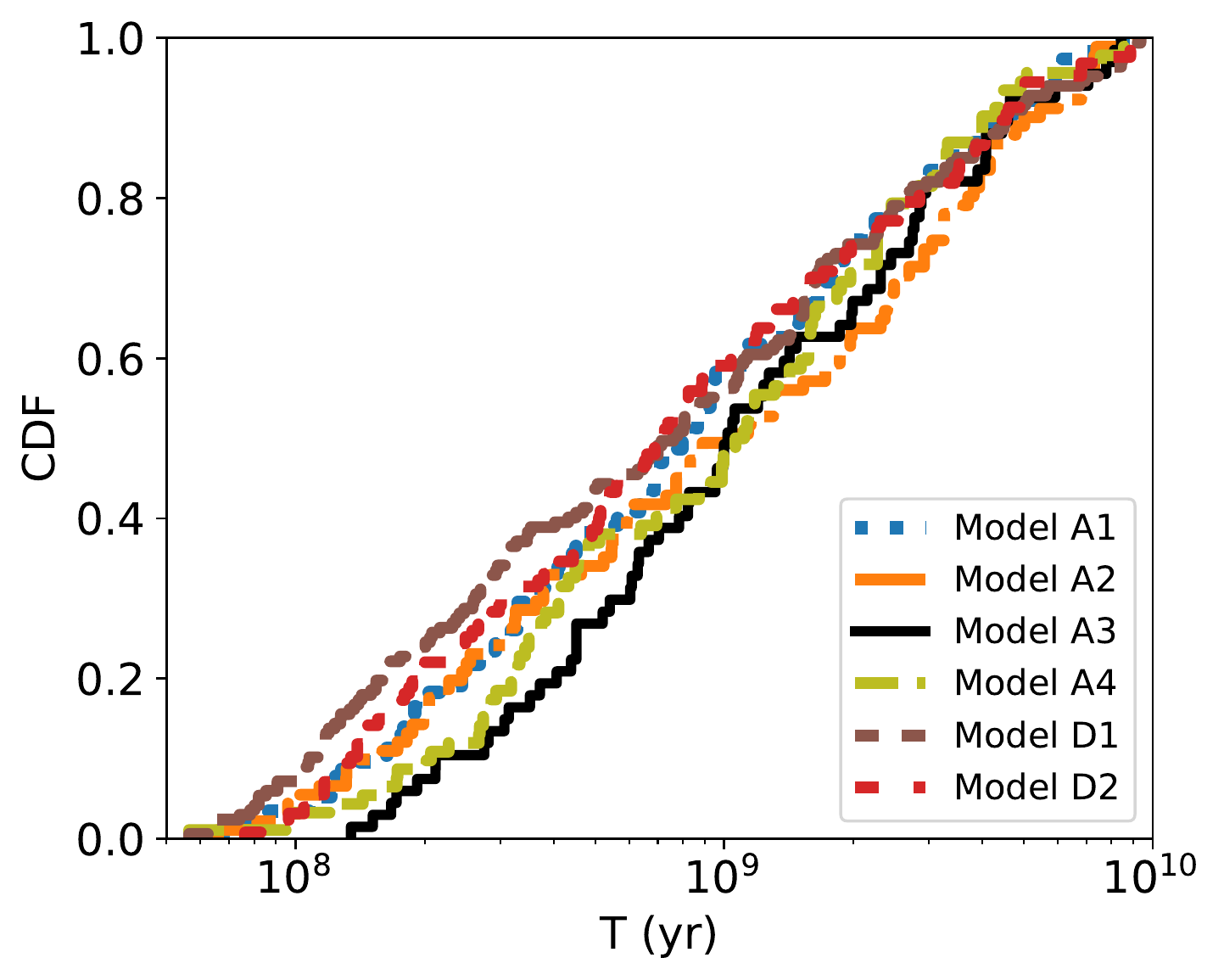}
\caption{WD-TDE time distribution (after the SN event) of BH-WD binaries in triples that lead to a WD-TDE for all models (see Tab.\ref{tab:models}).}
\label{fig:twdtde}
\end{figure}

Figure~\ref{fig:twdtde} reports the distribution of WD-TDE times for all models (see Tab.~\ref{tab:models}). The shape of these CDFs is quite universal and does not depend on the assumed value of the mean kick velocity for BHs and WDs nor on the initial distribution of semi-major axes and eccentricities. In order to compute the WD-TDE rate from BH-WD mergers in triples, we follow a similar calculation to that in \citet{sil17} and in \citet{fff2019}. We assume that the local star formation rate is $\eta_{\rm SFR} = 0.025 \msun$ Mpc$^{-3}$ yr$^{-1}$, thus the number of stars formed per unit mass, volume, and time is given by \citep{both2011},
\begin{align}
\dot{n}(m) &= \frac{\eta_{\rm SFR}\, f(m)}{\langle m\rangle}= 
\nonumber\\&=
5.2\times 10^6 \left(\frac{m}{\msun}\right)^{-2.3}\ \mathrm{M}_\odot^{-1}\ \mathrm{Gpc}^{-3}\mathrm{yr}^{-1}\ ,
\end{align}
where $\langle m\rangle = 0.38 \msun$ is the average stellar mass. Adopting a constant star-formation rate, the WD-TDE rate in triples is then,
\begin{align}
\mathcal{R}_\mathrm{WD-TDE} =&
\eta (1-\kappa) (1-\zeta) f_{\rm 3}\times\nonumber\\
&\times f_{\rm stable} f_{\rm WD-TDE}\int_{20\msun}^{150\msun} \dot{n}(m_1) dm_1
\nonumber\\=& 
7.4\times 10^4 \eta (1-\kappa) (1-\zeta) f_{\rm 3}\times\nonumber\\
&\times f_{\rm stable} f_{\rm WD-TDE}\ \mathrm{Gpc}^{-3}\ \mathrm{yr}^{-1}\ .
\end{align}
Here $f_{\rm 3}$ is the fraction of stars in triples, $f_{\rm stable}$ is the fraction of sampled systems that are stable after the SN events take place, and $f_{\rm WD-TDE}$ is the fraction of systems that produce a WD-TDE (see Tab.~\ref{tab:models}). The factor $\eta$ assures that, when sampling the mass ratio $q_{12}$ of the inner binary, the secondary ($1\msun \le m_2=q_{12}m_1\le 8\msun$) produces a WD,
\begin{equation}
    \eta = \frac{ \int_{0.01\msun}^{150\msun} d m_1 f_{\rm IMF}(m_1) \int_{{1\msun}/{m_1}}^{{{8\msun}/{m_1}}} dq_{12}  f_{q}(q_{12}) 
    }{
     \int_{20 \msun}^{150\msun} d m_1 f_{\rm IMF}(m_1) }\,,
\end{equation}
where $f_q(q_{12})$ is the mass ratio distribution of the inner binary. We get $\eta=0.21$ and $\eta=0.25$ for an uniform and log-uniform mass ratio distributions, respectively. The factors $\zeta$ and $\kappa$ take into account two main processes during the earlier evolution of the system which prohibit a WD TDE \citep{shapp2013}. The first comes from the fact that stellar triples can merge during their main sequence (MS) life before the primary forms a BH as a result of the LK dynamics, which we have not modeled here. To estimate $\zeta$, we conservatively consider that all stellar triples whose initial LK timescale is less than the lifetime of the primary star \citep[$\sim 7$ Myr;][]{iben91,hurley00,maeder09} in the inner binary merge as MS stars \citep{rodant2018}. We find that the fraction of these triples is $\zeta\sim 0.60$ on average, except for Model A3 and Model A4 where we find $\zeta\sim 0.35$. Furthermore, we check the fraction of systems that produce a stellar mean sequence TDE instead of a WD TDE, i.e. before the secondary star in the inner binary forms a WD. We estimate this fraction to be $\kappa\sim 0.15$ from the results of \citet{fff2019}.

In our calculations, we adopt for the triple fraction $f_{\rm 3}=0.25$ and $f_{\rm WD-TDE}\sim 0.21$ on average (see Tab.~\ref{tab:models}). The fraction of stable systems after the SNe depends on the value of $\sigbh$ and $\sigwd$, and on the details of the distributions of initial parameters. We report $f_{\rm stable}$ for all our models in Tab.~\ref{tab:models}. Using the minimum and maximum values of $f_{\rm stable}$ in Tab.~\ref{tab:models}, our final estimated WD-TDE rate is in the range,
\begin{equation}\label{eq:rate}
\mathcal{R}_\mathrm{WD-TDE}=(4.1\times 10^{-3}-4.8) \ \mathrm{Gpc}^{-3}\ \mathrm{yr}^{-1}\ .
\end{equation}
For a log-uniform distribution of mass ratios, we estimate a rate $\sim 1.5$ times larger. Considering the signal up to $z=0.1$, the WD-TDE rate becomes,
\begin{equation}\label{eq:Gamma1}
\Gamma_\mathrm{WD-TDE}(z\le 0.1)=(1.2\times 10^{-3}-1.4) \ \mathrm{yr}^{-1}\ .
\end{equation}

We can also estimate the WD-TDE rate in triples for a Milky Way-like galaxy. Assuming momentum-conserving natal kicks and a star formation rate of $1\msun\ \mathrm{yr}^{-1}$ \citep{licq2015}, we obtain
\begin{equation}\label{eq:Gammamw}
\Gamma_\mathrm{WD-TDE}^\mathrm{MW}=(4.8\times 10^{-11}-5.7\times 10^{-8}) \ \mathrm{yr}^{-1}\ .
\end{equation}

Finally, we note that we are not taking into consideration fallback in our calculations, whose effect would be to increase the WD-TDE rates for large $\sigbh$'s since it would give smaller natal kick velocities.

\subsection{The role of the mass loss prior to supernovae}

\begin{figure*} 
\centering
\includegraphics[scale=0.45]{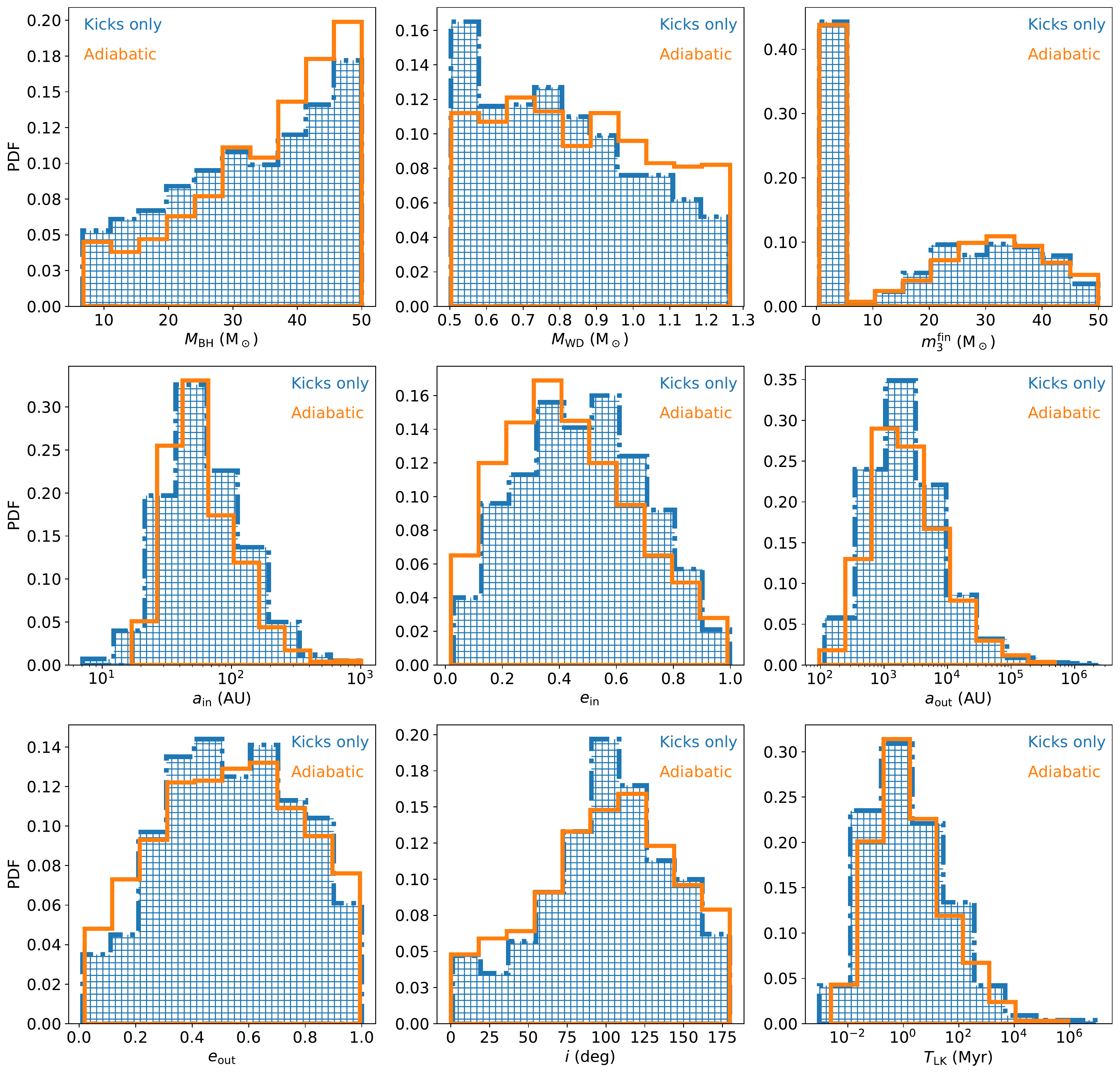}
\caption{Comparison of the distributions of masses, semi-major axes, eccentricities, inclinations, and LK timescales of stable systems, obtained in the case that only kick prescriptions are applied (blue histograms) and in the case that only adiabatic mass loss is applied (orange histograms) for Model A1 (see Tab.\ref{tab:models}).}
\label{fig:adiab}
\end{figure*}

In our simulations, we assume that the SN events take place instantaneously and do not simulate the systems during the main sequence lifetime of the progenitors, and the eventual mass loss prior to the SN explosion. Mass loss prior to SN events could change the binding energy of the triples, but also the effective mass ratios of the inner and outer components, semi-major axes, and eccentricities. This, along with kicks (taking place in wider systems), could affect the evolution of the triples consisting of an inner BH-WD binary.

To quantify the role of mass loss prior to SN, we now consider the impact of slow (adiabatic) and isotropic mass loss. Adiabatic mass loss drives the orbits of two objects to larger semimajor axes \citep[e.g.][]{kratt2012}
\begin{equation}
a_{\rm new}=\frac{m_{\rm old}}{m_{\rm new}}a_{\rm old}\ ,
\end{equation}
where $a_{\rm new}$ and $a_{\rm old}$ are the new and old semi-major axis, respectively, and $m_{\rm new}$ and $m_{\rm old}$ are the system mass after mass loss and before mass loss, respectively. In this approximation, the orbital eccentricity remains roughly constant. Since we assume in our calculations that all the mass loss takes place at the moment of SN, to bracket the uncertainties of our method, we apply the above prescription assuming that all the mass loss happens adiabatically for the three stars in our triple systems, with no mass loss at the moment of SN. In reality, mass loss will happen partially prior to the SN event and partially during the SN event itself.

We show in Figure~\ref{fig:adiab} the comparison of the distributions of masses, semi-major axes, eccentricities, inclinations, and LK timescales of stable systems, obtained for the case where only kick prescriptions are applied (blue histograms) and for the case where only adiabatic mass loss is applied for Model A1 (see Tab.\ref{tab:models}), using $1000$ realizations. We performed a two-sample KS test to assess quantitatively the statistical difference between the two populations. We find that the D-values of the parameters shown in Figure~\ref{fig:adiab} for the two populations are $\lesssim 0.043$, which corresponds to a $95\%$ confidence level, thus statistically consistent, except for $M_{\rm BH}$, $M_{\rm WD}$, and $e_{\rm in}$. We find that the adiabatic case favours higher BH and WD masses. We find similar results for the other models in Table~\ref{tab:models}.

We report in Table~\ref{tab:models} the fraction of stable systems assuming adiabatic mass loss. Compared to the fraction of stable systems assuming kicks only, we find that the fraction of stable triples that can lead to a BH-WD merger is increased by a factor of $\sim 3$--$20$ in the case that the mass loss is adiabatic only. Assuming a similar merger fraction as found in our simulations, this would even imply a slightly larger number of BH-WD mergers. We caution that both approaches are only approximate and a final answer is left to future studies (see Conclusions). However, this definitively shows that adiabatic mass loss prior to SNe can have a relevant effect on the stability and merger rates of BH-WD systems in triples.

\section{Electromagnetic counterparts of Black Hole-White Dwarf tidal disruption events}
\label{sect:emcount}

Previous works have considered possible electromagnetic signatures of the tidal disruption of a WD by a stellar-mass compact object, such as a NS or a BH \citep{Fryer+99,King+07,Metzger12,Fernandez&Metzger13,Margalit&Metzger16,Zenati+19,Fernandez+19}. As discussed, the WD is tidally disrupted once its orbital pericenter radius, $R_{\rm p}$, decreases below the tidal radius $R_{\rm T}$ (see Eq.~\ref{eqn:rtid}). The disruption of stars and planets was first discussed by \citet{Perets2016}, who termed these events micro-TDEs. More recently, these events have been discussed by \citet{fff2019} in triple systems and \citet{Kremer2019} in star clusters.

Tidal pinching of the WD and/or tidal tail intersection can in principle result in thermonuclear burning during the WD disruption process \citep{Luminet&Pichon89,Rosswog+09,Macleod+16,Kawana+18}, particularly for high penetration factors 
\be
\beta \equiv \frac{R_{\rm t}}{R_{\rm p}} = \left(\frac{R_{\rm p}}{R_{\rm WD}}\right)^{-1}\left(\frac{\mbh}{M_{\rm WD}}\right)^{1/3}\,.
\ee  
However, unlike the focus of the present paper, most of these works consider massive $\gtrsim 100 \msun$ black holes, for which high $\beta \gg 1$ and thus strong tidal compression is possible.\footnote{One exception is \citet{Kawana+18}, who in some cases obtain explosions with black holes of mass $10\msun$.} We do not generally expect significant nuclear burning during the disruption by lower-mass black holes.

\begin{figure} 
\centering
\includegraphics[scale=0.55]{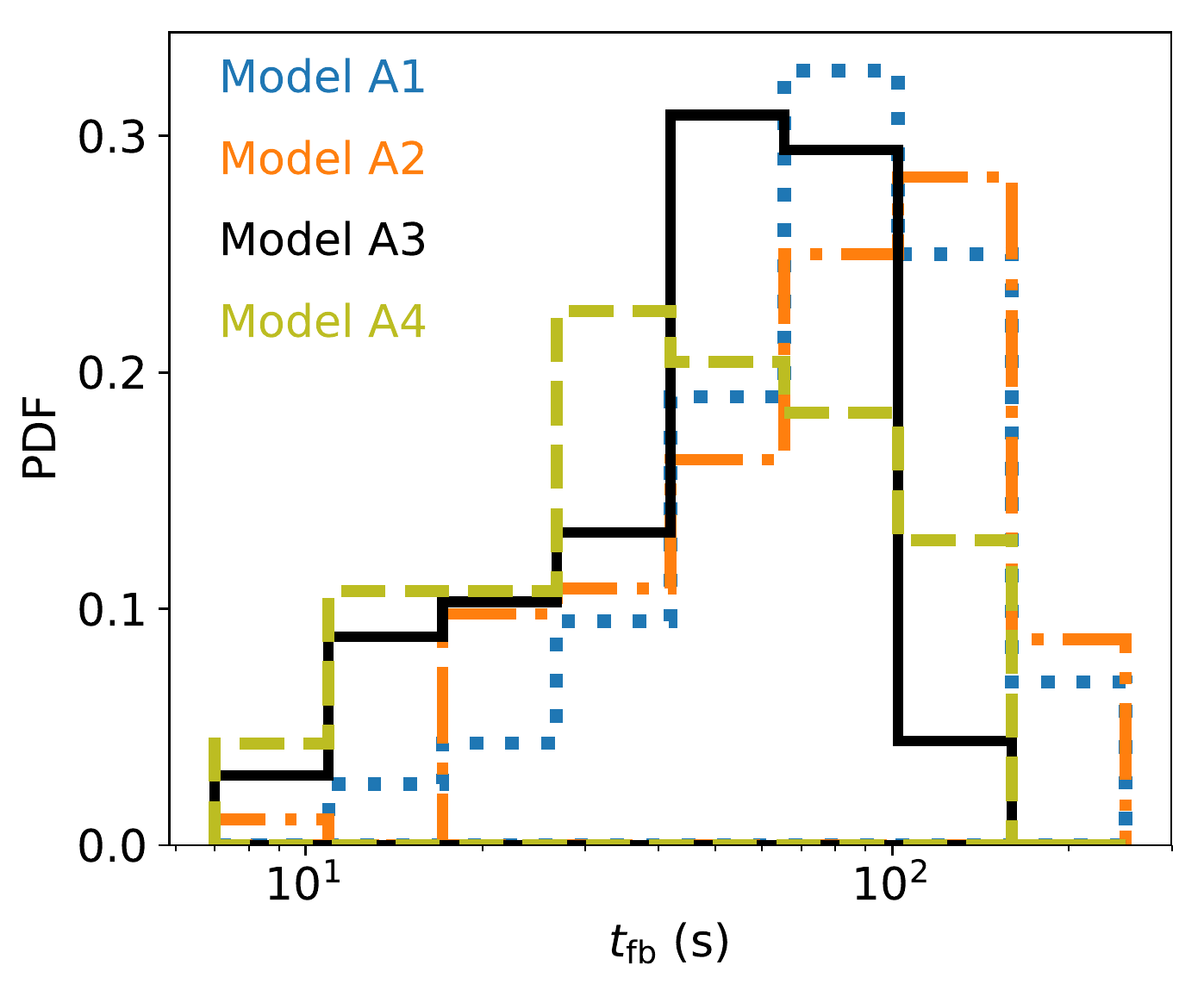}
\includegraphics[scale=0.55]{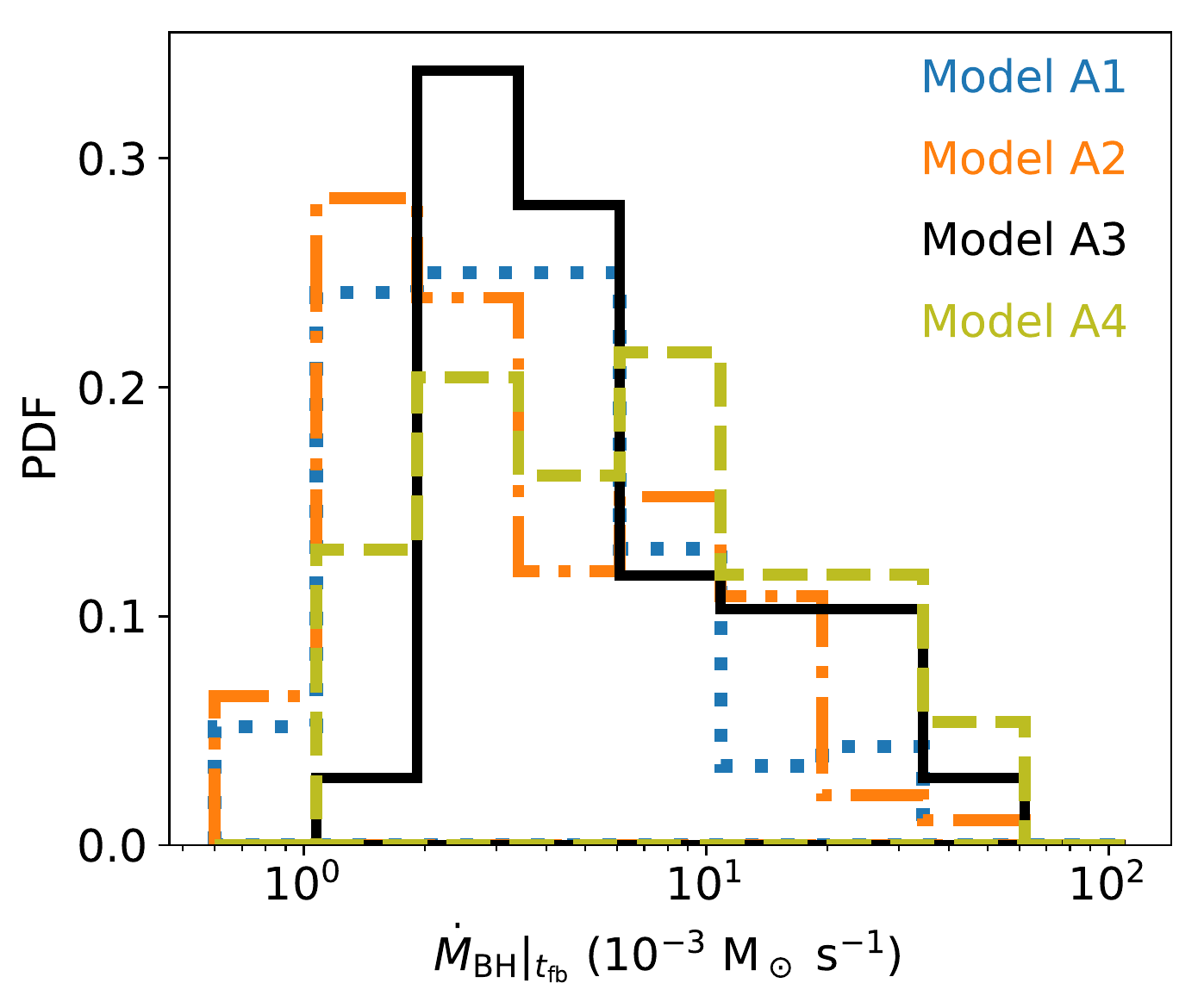}
\caption{Distributions of fall-back times (top) and accretion rates (bottom) for stellar BH-WD systems that lead to a merger for Models A1-A4.}
\label{fig:tfbmdot}
\end{figure}

The tidal disruption imparts a specific energy spread to the WD debris \citep{Rees88},
\be
\Delta E_{\rm t} \sim \frac{G\mbh R_{\rm WD}}{R_{\rm t}^{2}}.
\ee
This greatly exceeds the initial orbital binding energy of the WD, $E_{\rm orb} \sim G\mbh/a$, for initial WD semi-major axes obeying
\be\label{eq:at}
a \gg a_{\rm t} = \frac{R_{\rm t}^{2}}{R_{\rm WD}} = R_{\rm WD}\left(\frac{\mbh}{M_{\rm WD}}\right)^{2/3}\,.
\ee
The condition $\Delta E_{\rm t} \gg E_{\rm orb}$ is easily satisfied by the WD-TDEs in our population.  In this case, the half of the disrupted WD furthest from the BH at the time of disruption receives positive energy and is ejected promptly from the system.  The other half of the WD is tightly bound to the BH and returns to the tidal radius over a characteristic fall-back time corresponding to the orbital period of matter with binding energy $\Delta E_{\rm t} = G\mbh/a_{\rm t}$   (e.g.~\citealt{Stone+13})
\begin{eqnarray}\label{eq:tfb}
t_{\rm fb} &\sim& 2\pi \left(\frac{a_{\rm t}^{3}}{G\mbh}\right)^{1/2} \approx  2\pi \left(\frac{R_{\rm WD}^{3}\mbh}{GM_{\rm WD}^{2}}\right)^{1/2} \approx\nonumber\\
&\approx& 100\,\,{\rm s}\left(\frac{\mbh}{10\msun}\right)^{1/2}\left(\frac{M_{\rm WD}}{0.6\msun}\right)^{-1}\left(\frac{R_{\rm WD}}{10^{4}\,{\rm km}}\right)^{3/2}\,.
\end{eqnarray}
In the top panel of Figure~\ref{fig:tfbmdot}, we illustrate $t_{\rm fb}$ for Models A1-A4. 

Also note that we are justified in neglecting the influence of a binary companion on the dynamics of the mass fallback  \citep{Coughlin2017,liu2019}.  This is because the apocenter radii $\sim a$ of the bound debris $a \simeq 0.41 (M/10M_\odot)^{1/3} (T/1\,{\rm month})^{2/3}$~AU, 
where $T$ is the elapsed time
since the disruption, is much smaller than the separations of the outer companion of the systems considered here (Fig.~\ref{fig:ainaout}).  

For the bound fallback material to circularize and hence form an accretion disk, it
must lose a significant amount of energy. Circularization is believed to be aided by relativistic effects, since apsidal precession causes highly eccentric debris streams to
self-intersect (e.g. \citealt{Hayasaki2016,Sadowski2016,stone2019}). However, whether circularization can be fully realized before the end of the actual TDE still remains an issue of discussion (see e.g. \citealt{Piran2015}); in the case of stellar mass BHs, it is aided by the fact that the bound
debris are not highly eccentric \citep{Kremer2019}. Additionally, a large fraction
of the tidally disrupted material is expected to be flung out and become unbound as a
result of heating associated with inter-stream shocks
\citep{Ayal2000}. 

For the debris which remains bound, at times $t \gg t_{\rm fb}$, the fall-back rate obeys
\be
\dot{M}_{\rm fb} \approx \dot{M}_{\rm fb}|_{t_{\rm fb}}\left(\frac{t}{t_{\rm fb}}\right)^{-5/3},
\ee
where
\begin{align}
\dot{M}_{\rm fb}|_{t_{\rm fb}} \approx \frac{M_{\rm WD}}{3 t_{\rm fb}} \approx& \,2\times 10^{-3}\msun\,{\rm s^{-1}}\left(\frac{\mbh}{10\msun}\right)^{-1/2} \nonumber\\
&\times  \left(\frac{M_{\rm WD}}{0.6\msun}\right)^{2}\left(\frac{R_{\rm WD}}{10^{4}\,{\rm km}}\right)^{-3/2}
\label{eq:Mdotfb}
\end{align}
is the peak fall-back rate.  Once in a circular disk at $R_{\rm out} \sim 2 R_{\rm t}$, matter is fed onto the BH on the viscous timescale, 
\be
t_{\rm visc} \approx \frac{1}{\alpha}\left(\frac{R_{\rm out}^{3}}{G\mbh}\right)^{1/2}\left(\frac{H}{R_{\rm out}}\right)^{-2},
\ee
where $H/R_{\rm out} \sim 1$ is the aspect ratio of the disk and $\alpha$ its effective viscosity.  To the extent that $(\mbh/M_{\rm WD})^{1/2} \lesssim \alpha^{-1}$, the viscous timescale is generally longer than the fall-back time.  However, for simplicity we adopt Eq.~\eqref{eq:Mdotfb} for the BH accretion rate hereafter (though note that the true accretion rate could be smaller if $t_{\rm visc} \gg t_{\rm fb}$).

As matter accretes deeper into the potential well approaching the BH, the increasingly high densities and temperatures of the accretion flow will burn the WD material into increasingly heavy elements at sequentially smaller radii, generating an onion-skin like radial structure to the disk composition \citep{Metzger12}.  

Given the very high accretion rates, photons are trapped in the accretion flow and radiative cooling is inefficient.  Under these conditions, powerful disk winds driven by the released accretion energy are likely to carry away most of the accreting material before it reaches the central BH (e.g.~\citealt{Narayan&Yi95,Blandford1999b}).  Axisymmetric hydrodynamical $\alpha$-viscosity simulations by \citet{Fernandez+19} find that the accretion rate $\dot{M}_{\rm BH}$ which ultimately reaches the innermost stable circular orbit of the central BH ($R_{\rm BH} = 6R_{\rm g}$, where $R_{\rm g} = GM_{\rm BH}/c^{2}$) is reduced from the outer feeding rate $\dot{M}_{\rm fb}$ according to
\begin{eqnarray}
\dot{M}_{\rm BH}|_{t_{\rm fb}} &\approx & \dot{M}_{\rm fb}|_{t_{\rm fb}}\left(\frac{R_{\rm BH}}{2R_{\rm t}}\right)^{p} \approx  2.6\times 10^{-5}\msun\,{\rm s^{-1}}\times\nonumber\\
&\times & \left(\frac{\mbh}{10\msun}\right)^{-0.03}\left(\frac{M_{\rm WD}}{0.6\msun}\right)^{2.23}\left(\frac{R_{\rm WD}}{10^{4}\,{\rm km}}\right)^{-2.2}
\label{eq:MdotBH}
\end{eqnarray}
where $p \approx 0.7$.  The $\sim 99\%$ of the matter not accreted by the BH is ejected in a wind (e.g.~\citealt{Margalit&Metzger16,Fernandez+19}).  In the bottom panel of Figure~\ref{fig:tfbmdot}, we illustrate $\dot{M}_{\rm BH}|_{t_{\rm fb}}$ for Models A1-A4. The inner parts of the accretion flow (near the central BH) could generate a relativistic jet similar to those which give rise to gamma-ray bursts (e.g., \citealt{Fryer+99,King+07}).  From Fig.~\ref{fig:tfbmdot},  we predict peak accretion rates of $\sim 10^{-3}-10^{-2}\msun$ s$^{-1}$, with a typical timescale varying between a few tens and a few hundreds of seconds (models A1 and A2 have especially long timescales).  Assuming a jet launching efficiency of $\epsilon_{\rm j} \lesssim 0.1$, the peak jet power could therefore be $\epsilon_{\rm j}\dot{M}c^{2} \lesssim 10^{50}-10^{51}$ erg s$^{-1}$.  Internal dissipation (e.g. shocks or magnetic reconnection) within such a jet could plausibly give rise to a short-lived ($\sim$ several tens to several hundreds of seconds) gamma-ray or X-ray transient with a luminosity $10^{49}-10^{50}$~erg~s$^{-1}$.

Have such events already been observed?
The distribution of accretion rates and durations largely falls within 
the range observed for the canonical long GRBs, with the tail of the distribution approaching the timescales of 
 a sub-class with similar characteristics, but longer duration,
known as the 'ultra-long' GRBs (ULGRBs) \citep{Levan+14}. 
It remains debated whether ULGRBs are simply the longest lasting members of a single, continuous LGRB population \citep{Zhang+14} or whether they represent a distinct class with potentially different progenitors \citep{Levan+14}.  
Either way, an important observational feature of the TD events discussed here would be the lack of signatures of a core-collapse supernova in association with a GRB-like type event. Wolf-Rayet stars are believed to be the progenitor stars of the canonical LGRBs \citep{Woosley2006}, while a working model for the ULGRBs is the collapse of a blue supergiant star (\citealt{Perna2018b}; though competing models exist involving millisecond magnetar engines, e.g.~\citealt{Greinder+15,Metzger+15}). Among the events observed so far, an interesting one is GRB060614, with a duration of 102~s. At a redshift of $z=0.125$, its associated core-collapse SN should have been detected, but it was not, calling for the possibility of a new $\gamma$-ray burst classification \citep{Gehrels2006}, which \citet{King+07} suggested might be indicative of a WD-NS merger. Future events of this kind will hence deserve special attention. 

However, while not accompanied by canonical core-collapse SNe, 
WD-BH mergers may be accompanied by fast-evolving supernova-like transients.  As mentioned above, half of the white dwarf is unbound promptly during the tidal disruption process at a characteristic velocity $v_{\rm t} \sim (G\mbh^{1/3}M_{\rm WD}^{2/3}/R_{\rm WD})^{1/2} \approx 4.4\times 10^{3}$ km s$^{-1}$.  Due to outflows from the accretion disk, the majority of the bound half of the WD will also be ejected, with a range of velocities $\sim 10^{4}-10^{5}$ km s$^{-1}$ \citep{Metzger12,Margalit&Metzger16,Fernandez+19}.  Due to the low ejecta mass $\sim M_{\rm WD} \lesssim 1 \msun$, any thermal transient would be expected to peak much faster than normal supernovae, e.g. on a timescale of $t_{\rm sn} \sim$ days instead of weeks.    

What source of luminosity would power the supernova-like emission?  Although little radioactive $^{56}$Ni is likely to be produced during the tidal compression, a small quantity of $^{56}$Ni is produced by the inner regions of the accretion flow \citep{Metzger12}.  \citet{Fernandez+19} predict that the accretion flows generated by the merger of quasi-circular BH-WD systems will generate $\sim 10^{-3}-10^{-2}\msun$ in ejected $^{56}$Ni (see their Table 2), in which case the resulting thermonuclear supernovae would peak at a luminosity $\sim 10^{40}-10^{41}$ erg s$^{-1}$, i.e.~10-100 times less luminous than normal Type Ia supernovae.  

The luminosity of the thermonuclear supernova could be substantially boosted if the ejecta is heated by ongoing outflows from the central engine (e.g.~\citealt{Dexter&Kasen13}).  In particular, if 10\% of the accretion power reaching the BH goes into powering the supernova luminosity through accretion-disk winds, then from equations (\ref{eq:Mdotfb}) and (\ref{eq:MdotBH}), we see that the peak luminosity could reach
\be
L_{\rm pk} \approx 0.1\dot{M}_{\rm BH}|_{t_{\rm fb}}c^{2}\left(\frac{t_{\rm sn}}{t_{\rm fb}}\right)^{-5/3} \sim 10^{43}-10^{44}{\rm erg\,s^{-1}}
\ee
for characteristic parameters, e.g. $t_{\rm sn} \sim 1$ d.  Consistent with this, numerical simulations of the  radiation-hydrodynamic evolution of disk winds by \citet{Kremer2019} showed that it can produce optical transients with luminosities $\sim 10^{41}-10^{44}$~erg~s$^{-1}$, on timescales varying from about a day to about a month.

A number of fast-evolving blue supernovae with luminosities in this range have been discovered in recent years \citep{Drout+14}.  However, the total estimated rate of this (highly heterogeneous) population of transients $\approx 4800-8000$ Gpc$^{-3}$ yr$^{-1}$  exceeds the rate of BH-WD mergers found in this paper, hence suggesting another progenitor origin for the bulk of this population. Furthermore, the closest example yet discovered (AT2018cow; e.g.~\citealt{Prentice+18}) showed evidence for hydrogen in the spectrum, ruling out a BH-WD origin.

\section{Gravitation wave counterparts}
 
It is well known that the extreme mass ratio inspiral of a WD into an IMBH is a target for multimessenger observations including GW detections with \textit{LISA} \citep[e.g.][]{HilsBender1995,Kobayashi+2004,DaiBlandford2013,Eracleous+2019}. Stellar-mass BH--WD and NS--WD inspirals could also be observed coincidentally in GWs with \textit{LISA} up to the point of disruption. The GW frequency at disruption is\footnote{This expression corresponds to circular orbits, but the peak GW frequency at disruption is similar for arbitrary eccentricities $0\leq e \leq 1$ to within $20\%$.}
\begin{align}\label{eq:fgw}
    f_{\rm GW} &= 
       \frac{G^{1/2}  (M_{\rm BH}+M_{\rm WD})^{1/2} }{\pi R_{\rm t}^{3/2}}
    \nonumber\\&= 
       0.09\,{\rm Hz}\,\left(1+\frac{M_{\rm WD}}{M_{\rm BH}}\right)
       M_{\rm WD,0.6\msun}^{1/2} 
       R_{\rm WD,10^4\,{\rm km}}^{-3/2}
\end{align}
where we introduced the abbreviated notation $X_{,a} = X/a$. Note that for $R_{\rm WD}\propto M_{\rm WD}^{-1/3}$ and $M_{\rm WD}\ll 1.44\msun$ (see Eq.~\ref{eq:Rwd}), the GW frequency at disruption follows $f_{\rm GW}\propto M_{\rm WD}$ and is independent of $M_{\rm BH}$ to leading order.

\begin{figure} 
\centering
\includegraphics[scale=0.55]{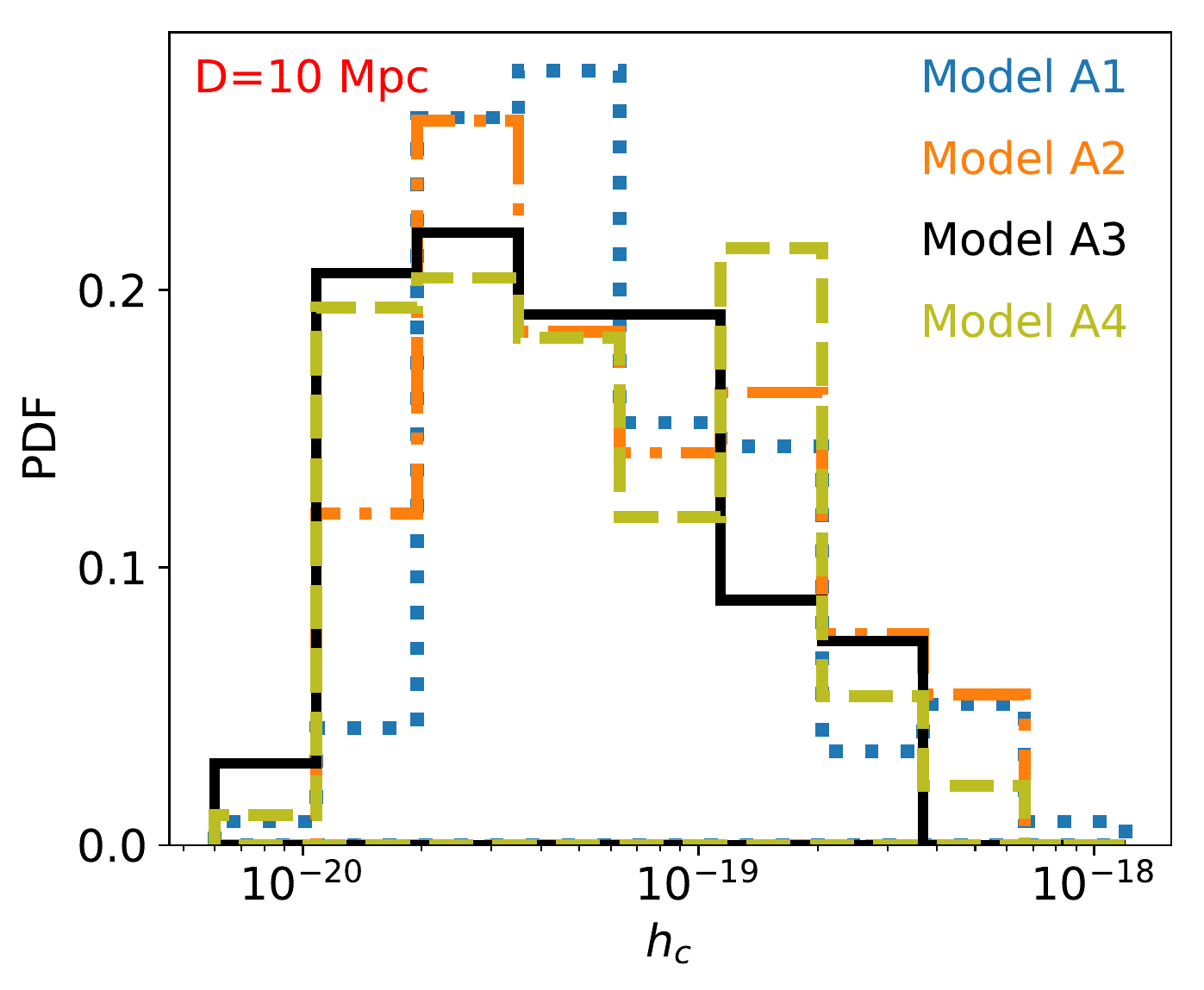}
\caption{Distributions of the total characteristic GW strain $h_{\rm c}$ for stellar-mass BH-WD systems that lead to a merger at a distance of 10 Mpc for Models A1-A4. For comparison, note that the characteristic \textit{LISA} noise amplitude is approximately
$8\times 10^{-21}$ at 0.09 Hz corresponding to the TDE.
}
\label{fig:straingw}
\end{figure}

The total characteristic GW strain for observing the GWs for a duration $T$ averaged over binary and detector orientation is approximately \citep[see Eq. 27 in][]{Robson+2019}
\begin{align}
    h_{\rm c} &= \frac{8}{\sqrt{5}} \frac{G^2}{c^4}\frac{M_{\rm BH}M_{\rm WD}}{R_t D} \left(T f_{\rm GW}\right)^{1/2}  = 2.0\times 10^{-20}\nonumber\\
    &\quad\times T_{4\rm yr}^{0.5} D_{10\, \rm Mpc}^{-1} 
    M_{\rm BH,10\msun}^{0.66} M_{\rm WD,0.6 \msun}^{1.58}
    R_{\rm WD,10^4\,{\rm km}}^{-1.75}
\end{align}
In Fig.~\ref{fig:straingw} we show the distributions of the total characteristic GW strain for stellar-mass BH-WD systems that lead to a merger in Models A1-A4, at a distance of $D=10$ Mpc. Note that the characteristic noise amplitude for \textit{LISA} is $8\times 10^{-21}$ at 0.09 Hz \citep[see Fig. 6 in][]{Robson+2019}.
We follow \citet{Robson+2019} to calculate the signal to noise ratio (SNR) of detecting the GWs for a WD-BH inspiral using the current design of \textit{LISA} with arm length of $2.5\times 10^6\,$km. For a 4 year observation, we find that the binary orientation averaged SNR is higher than 8 up to 10 Mpc for $M_{\rm BH}=40\msun$, $M_{\rm WD}=1.0\,\msun$. For optimal (face-on) binary orientation, the detection distance is a factor 2.5 larger. We conclude that the GW observations of WD mergers with stellar-mass BHs or NSs will be limited to the local Universe within $\sim 25$ Mpc. 

The detection volume is $V_{\rm GW} \propto h_c^3/[f S_n(f)]^{1/2}$, where $S_n$ is the noise power spectral density, which scales as $S_n(f)\propto f^2$ for $f\gtrsim 0.02\,$Hz. Thus $V_{\rm GW}\propto h_c^3 f_{GW}^{-3/2}\propto M_{\rm BH}^2 M_{\rm WD}^{4} R_{\rm WD}^{-3} T^{1.5}$. The GW-observed TDE rate is strongly biased towards higher BH mass and higher WD masses.

If a NS-WD, BH-WD, or IMBH-WD TDE happens within the detectable \textit{LISA} volume, the GW measurements can be used to determine the parameters $f_{\rm GW}$ given by Eq.~\eqref{eq:fgw}, the chirp mass of the binary $\mathcal{M}=M_{\rm BH}^{3/5}M_{\rm WD}^{3/5}/(M_{\rm BH}+M_{\rm WD})^{1/5}$, and the distance to the source $D$ independently of electromagnetic observations. Coincident GW detections may help to secure the identification of the electromagnetic counterpart. The joint multimessenger analysis of the merger of a WD  with a stellar BH, NS, or IMBH offers to gain a more accurate  understanding of these astrophysical sources. Our predicted rate implies that the chance of coincident EM/GW detections of tidal disruption events in stellar triple systems is very small as the rate is  $\sim 10^{-6}-10^{-9}$ yr$^{-1}$ within $10$ Mpc. Most WD/BH binaries detectable with LISA are expected to be far from merger.

Note that BH--WD and NS--WD binaries typically emit GWs in the \textit{LISA} band for thousands of years before merger. We find that the peak signal to noise ratio (SNR) corresponds to $f\sim 0.03\,\rm Hz$ and the SNR decreases slowly for higher $f$. For a fixed observation time, the SNR varies by less than a factor 2.5 between $0.005$Hz and the point of tidal disruption. For circular BH-WD binaries, the GW frequency arrives at  $0.005$Hz at $10^4\,{\rm yr} M_{\rm WD,0.6\msun}^{-1}M_{\rm BH,10\msun}^{-5/3}$ before tidal disruption. Since the number of systems at a given frequency $f$ scales with the residence time $N \propto t_{\rm res} = f/\dot{f} \propto f^{-8/3}$, most WD/BH binaries detected through GWs will be far from disruption. Furthermore, LK oscillations induced by a triple companion leaves a time-dependent imprint on the GW spectrum of the inner binary \citep{Randall2019,Hoang+2019}.
Thus, \textit{LISA} observations of GWs emitted by binaries in the Galaxy may be used directly to constrain the expected TDE rate from BH--WD and NS--WD TDE systems in the Universe.

\section{Conclusions}
\label{sect:conc}

The mergers of binaries comprised of two compact objects can produce diverse explosive transient events, such as GW chirps, Type Ia and GRBs.  Though they have received comparatively less attention in the literature, the mergers of NS-WD and BH-WD binaries are expected to generate transients if the WD approaches the NS or BH close enough to be disrupted in a WD-TDE.

This paper explores a new triple channel for WD-BH mergers driven by the joint effect of GW emission and the LK mechanism.  We explore the sensitivity of our results to different assumptions for the distributions of natal kick velocities imparted to the BH and the WD, the semi-major axes and eccentricities of the triple and the initial stellar masses.  We estimate the rate of WD-TDEs in triples to be in the range $1.2\times 10^{-3}-1.4$ Gpc$^{-3}$ yr$^{-1}$ for $z\leq 0.1$, under the assumption of momentum-conserving natal kicks. Compared to stellar TDEs in triples, WD-TDEs are therefore a factor of $\sim 3$--$30$ rarer. Moreover, we have found that the fraction of stable triples that can lead to a BH-WD merger is enhanced by a factor of $\sim 3$--$20$ in the case that the mass loss is adiabatic only. Assuming a similar merger fraction as found in our simulations, this would even imply a slightly larger number of BH-WD mergers, assessing the relevance of adiabatic mass loss prior to SNe in triples.

In our simulations we check that the triple systems remain stable after each SN event. Systems that become unstable may still merge, but they are not taken into account in our results. Moreover, we are assuming that the SN events take place instantaneously and do not simulate the systems during the main sequence lifetime of the progenitors. This and the details of the specific evolutionary paths, which depend on stellar winds, metallicity and rotation, of the stellar progenitors could reduce the available parameter space for BH-WD mergers \citep{shapp2013}. Systems that experience significant LK oscillations on the main sequence will frequently be driven to merge prematurely during this phase, assuming the LK process is not damped by GR or tidal precession in the inner binary. Such systems will typically be those triples with high birth inclinations, even though this is not a necessary requirement in cases where the octupole term or chaotic non-secular dynamics become important. 

Therefore, to avoid counting systems that would undergo main sequence mergers prior to the start of our simulations, we have conservatively neglected the contribution of stable triple systems which have LK timescales shorter than the lifetime of the progenitor stars in the inner binary (hence our inferred merger rates should be interpreted as lower limits on the true rate). The systems that we account for, which can possibly lead to a merger, are therefore the triples that were born typically with moderate inclinations, which only become highly inclined due to the BH kick.  Moreover, we also tried to correct for mergers which occur before the inner binary secondary evolves to the WD phase by accounting for the progenitors that could lead to a stellar TDE, rather than a WD-TDE, thus removing another portion of triples \citep{shapp2013,toonen2016,toonen2018,fff2019}. 

The situation becomes even more complicated if episodic mass loss occurs due to eccentric Roche-lobe overflow and/or if common evolutionary phases in the triple are taken into account.  However, these are not modeled in a self-consistent way in triple systems.  This is  because of a possibly complex interplay between these effects and LK evolution during the main sequence lifetime of the progenitors \citep{leigh16,rosa2019,hamd2019}.

Accretion of the bound debris onto the BH following a BH-WD TDE could power a relativistic jet, generating a burst of high energy X-ray or gamma-ray emission with a duration similar to a long GRB.  The heating of white dwarf debris (unbound during the tidal disruption event or in outflows from the accretion disk) by radioactivity or winds from the accretion disk, could generate a rapidly-evolving supernova-like optical transient.  Such peculiar transients from BH-NS mergers might be observable by high energy satellites or upcoming time-domain optical surveys, such as LSST.  The characterisation of WD-TDE events and their distributions is therefore a fundamental step in ultimately being able to identify them among the myriad of other cataclysmic events. 

Stellar mass BH-WD, NS-WD, and IMBH-WD, binaries may also be detected in GWs using \textit{LISA} up to the point of TDE. \textit{LISA} may also provide an accurate determination of the TDE rates from triples by observing systems thousands of years before merger in the Galaxy. Multimessenger studies of WD TDEs by stellar-mass BHs or NSs will be limited by the \textit{LISA} detection range of $\sim$10 Mpc. 

The future discovery of a population of WD-TDE could be used to study the demographics of BHs in nearby galaxies and to place constraints on the distributions of natal kicks at BH birth in a complementary way to what now probed by LIGO from BH-BH mergers.

\section*{Acknowledgements}

GF thanks Seppo Mikkola for helpful discussions on the use of the code \textsc{archain}. We thank our referee, Hagai Perets, for a constructive and stimulating report. GF acknowledges support from a CIERA postdoctoral fellowship at Northwestern University. RP acknowledges support by NSF award AST-1616157. The Center for Computational Astrophysics at the Flatiron Institute is supported by the Simons Foundation.  NWCL gratefully acknowledges a Fondecyt Iniciacion grant (\#11180005). This work received funding from the European Research Council (ERC) under the European Union’s Horizon 2020 Programme for Research and Innovation ERC-2014-STG under grant agreement No. 638435 (GalNUC) and from the Hungarian National Research, Development, and Innovation Office under grant NKFIH KH-125675 (to BK).  BDM acknowledges support by NASA through the Astrophysics Theory Program (grant number NNX17AK43G) and by the Simons Foundation (grant number 606260).

\bibliographystyle{mn2e}
\bibliography{refs}

\end{document}